\documentclass[reprint,
 amsmath,amssymb,
 aps,
superscriptaddress]{revtex4-2}
\usepackage{bm, color}

\usepackage[colorlinks = true,
            allcolors = blue]{hyperref}
\usepackage{placeins}
\usepackage{float}
\usepackage{hyperref}
\usepackage{braket}
\usepackage{xcolor}
\usepackage{algpseudocode}

\usepackage{graphicx}
\usepackage{dcolumn}
\usepackage{bm}
\usepackage{natbib}
\usepackage{mhchem}

\usepackage{siunitx}
\usepackage{color,soul}

\begin{document}

\preprint{APS/123-QED}
\title{Consistent inclusion of triple substitutions within a coupled cluster based static quantum embedding theory}

\author{Avijit Shee}
\affiliation{Department of Chemistry, University of California, Berkeley, CA 94720, USA}
\author{Fabian M. Faulstich}
\affiliation{Department of Mathematics, Rensselaer Polytechnic Institute, Troy, NY 12180, USA}
\author{K. Birgitta Whaley}
\affiliation{Department of Chemistry, University of California, Berkeley, CA 94720, USA}
\affiliation{Berkeley Center for Quantum Information and Computation, Berkeley, CA 94720, USA}
\author{Lin Lin}
\affiliation{Department of Mathematics, University of California, Berkeley, CA 94720, USA}
\affiliation{Applied Mathematics and Computational Research Division, Lawrence Berkeley National Laboratory, Berkeley, CA 94720, USA}
\author{Martin Head-Gordon}
\affiliation{Department of Chemistry, University of California, Berkeley, CA 94720, USA}


\date{\today}
\begin{abstract}

We have previously proposed the MPCC static embedding framework for quantum chemistry that self-consistently couples a high-level coupled cluster (CC) treatment of the fragment (active region) with a lower level, M{\o}ller-Plesset perturbatation treatment of the environment. Our initial implementation was limited to single and double (SD) substitutions, with CCSD for the fragment, and first-order perturbative SD amplitudes for the environment. Here, we extend the MPCC embedding treatment to triple substitutions, which is essential for achieving chemical accuracy in energy differences. To this end, we employ a CCSDT solver for the fragment subsystem. For the environment subsystem, we construct a perturbative estimate of the triples amplitudes, explicitly accounting for feedback from all fragment amplitudes. The resulting approach is denoted MPCCSDT(pt). We further introduce a more complete formulation in which feedback from the environment amplitudes to the fragment amplitudes is also included. This scheme involves an iterative treatment of the environment triples amplitudes and is denoted MPCCSDT(it). In addition, we assess the accuracy of the previously proposed low-level method by introducing a modified low-level approach that incorporates a lowest-order treatment of selected long-range effects, including spin fluctuations and charge polarization. All resulting approaches may be viewed as post-CCSD(T) methods. We therefore consider test cases for which CCSD(T) exhibits substantial deviations from CCSDT. These include (i) single- and triple-bond stretching in \ce{F_2} and \ce{N_2}, (ii) bond dissociation energies of selected molecules from the W4-11 dataset, and (iii) total atomization energies of transition metal hydrides. Our results demonstrate that inclusion of triples amplitudes at the fragment level alone is insufficient; a perturbative treatment of the environment triples amplitudes is required. For many energy-difference applications, feedback from the environment triples amplitudes to the fragment amplitudes, is not essential, but it does play a role in the very challenging CoH and FeH molecules. A very interesting finding from our study is that in some challenging cases we need an improved (second-order) perturbative method for the SD amplitudes, going beyond the first-order one used in our earlier work. Considering both cost and accuracy, the MP2CCSDT(pt) model is the most promising for future applications among the candidates considered here.

\end{abstract}

\maketitle

\section{\label{Intro}Introduction}

Wavefunction-based approaches to quantum chemistry are most commonly based on a single determinant reference, such as the mean-field Hartree-Fock wavefunction. The electron correlation energy \cite{Lowdin01011956}, $E_\mathrm{corr} = E-E_\mathrm{HF}$, is only $\sim 1$\% of the total energy, but is critical for calculating reliable energy differences because $E_\mathrm{corr}$ changes in most chemical transformations (albeit with important exceptions such as isodesmic reactions \cite{isodesmic_pople_jacs1970} and the like) as bonds are made and broken. Treating electron correlation at just the level of singles and doubles (SD) is also not adequate to achieve quantitative accuracy in energy differences, as has been known since the first implementations of triples (T) as part of fourth order M{\o}ller-Plesset (MP) perturbation theory\cite{MP4_FRISCH_CPL1980}.

Coupled cluster (CC) \cite{Cizek_CC, Paldus_Cizek} theory has become the predominant approach to electron correlation, as truncation of the amplitudes at any level of excitation (CCSD, CCSDT, etc) leaves polynomial-scaling equations for the amplitudes and guarantees size-consistent energies. The comparative compute costs and storage requirements for low-order CC theory, as well as for low-order MP theory are summarized in Table \ref{tab:cc_summary}. Based on the very large gap in storage and compute requirements between CCSD and CCSDT, the perturbative treatment of triples, CCSD(T) \cite{CCSD_TMHGCPL89}, has become the most popular way of including the triples. For most cases where the HF reference is a qualitatively good starting point, CCSD(T) is very successful, to the extent that it has become known as the ``gold standard'' of quantum chemistry. Furthermore, CCSD(T) has storage requirements that do not exceed CCSD, and its compute effort scales as non-iterative $\mathcal{O}(O^3V^4)$ versus iterative $\mathcal{O}(O^3V^5)$ for CCSDT.

\begin{table*}[t]
\centering
\begin{tabular}{lcccc}
\hline
\textbf{Method} & \multicolumn{2}{c}{\textbf{Iterative}} & \multicolumn{2}{c}{\textbf{Perturbative}} \\
\hline
 & Compute & Memory & Compute & Memory \\
\hline
MP2       & - & - & $\mathcal{O}(O^2V^2X)$ & $\mathcal{O} (O N X)$   \\
MP3       & - & - & $\mathcal{O}(O^2V^4X)$ & $\mathcal{O} (O N X)$   \\
CCSD      & $\mathcal{O}(O^2V^4)$ & $\mathcal{O}(O^2V^2)$ & - & -   \\
CCSD(T)   & $\mathcal{O}(O^2V^4)$ & $\mathcal{O}(O^2V^2)$ & $\mathcal{O}(O^3V^4)$ & -  \\
CCSDT   & $\mathcal{O}(O^3V^5)$ & $\mathcal{O}(O^3V^3)$ & $\mathcal{O}(O^3V^4)$ & -  \\
\hline
\end{tabular}
\caption{Compute and memory scaling requirements with problem size for conventional coupled cluster theory through triple substitutions and low-order M{\o}ller-Plesset perturbation theory. $O, V, X$ denote the size of the occupied space, the virtual space, and the auxiliary basis set, respectively.}
\label{tab:cc_summary}
\end{table*}

However, there are some regimes in which CCSD(T) demonstrably fails. Most clear-cut are the cases where triple substitutions are not a perturbative correction to the wavefunction, as assumed in CCSD(T), but are instead quite large. Such cases are typical of molecules or systems that exhibit strong electron correlations, such as partially recoupled unpaired electrons. For example, in \ce{Cr2}, 2 $^7$Cr atoms are recoupled to $^1$\ce{Cr2}, but the binding energy of 137 kJ/mol is weaker than a typical single bond rather than a hextuple bond\cite{Cr_dimer_expt_JPC93}. Other transition metal diatomics with partial spin recoupling are also very challenging for CCSD(T) \cite{Peterson_JCP2017}. Metalloenzymes with multiple metal centers are also considered to be strongly correlated molecules. Another class of failure of CCSD(T) is when the HF orbitals exhibit significant symmetry breaking (such as spin contamination) which is not properly recovered in its triples correction. If the symmetry-breaking is artificial (rather than a consequence of strong correlation) then use of alternative orbitals such as from Kohn-Sham density functional theory (DFT) can be helpful \cite{CCSDT_DFT_orbitals_MHG_JCTC2021, DFT_orbitals_CCSDT_Peterson_JCTC2016}. Finally, even in the intermolecular interactions between large $\pi$-electron molecules such as the so-called Buckycatcher complex, there is unresolved debate over whether CCSD(T) is adequate or not \cite{Tkatchenko_Nagy_NatComm2021,  Schafer_Gruneis_PhysRevLett2023,schafer_gruneis_NatComm2025}.

The strategy used in methods like CCSD(T) and its higher order analogs such as CCSDT(Q) \cite{Bartlett_JCP98} is an all-or-nothing choice between which amplitudes are treated self-consistently by CC theory, and which are approximated perturbatively. In other words, in CCSD(T), the SD amplitudes are iterated, and the T amplitudes are treated perturbatively. Indeed in CCSD and CCSDT all amplitudes are iterated and none are treated perturbatively. However, not all amplitudes at a given level of substitution are equally significant. Typically only a small fraction are large (corresponding to excitations within the valence space), while others that involve the very high energy price of promotions in the principal quantum number to access higher virtual levels are typically small. Therefore, it is reasonable to seek methods that do not make the ``all or none'' decision for a given class of amplitudes, but instead partition them into a ``primary set'' that should be iterated, and a ``secondary set'' that can be perturbatively approximated.

To make such an approach well-defined, rather than depending upon some arbitrary threshold for switching between primary and secondary, it is best to introduce a well-defined partition of the 1-particle orbitals into a primary and secondary set. The primary set could be a small basis (even a valence minimal basis) and the secondary set could be the missing orbitals (core and higher virtuals) necessary to comprise the target extended basis. This is reminiscent of dual basis methods for HF and DFT calculations\cite{steele2006dual, steele2007dual, steele2009non, liang2004approaching}, but carries the concept much further. In fact it is a particular type of quantum mechanical/quantum mechanical (QM/QM) embedding theory, where the question is how best to couple the primary amplitudes (that involve substitutions \textit{entirely within} the primary orbital set) and the secondary amplitudes (that involve substitutions that at least \textit{partly} involve the secondary orbital set).

A rigorous (QM/QM) embedding framework was first formulated in the context of Green’s function–based many-body theory by Georges \textit{et al.} \cite{DMFT_infinite_dim_Georges_Kotliar_1992,Georges96} in their dynamical mean field theory (DMFT) work. The DMFT formulation utilizes frequency-dependent/dynamical quantities such as Green's functions and the self-energy for the embedding construction. However, if static observables, such as total energy, dipole moment, spectroscopic parameters, etc. are the target quantities, a simpler static formulation is also possible. Density matrix embedding theory (DMET)\cite{Knizia12_DMET,Knizia13_DMET} is a prominent example of this direction. 

A crucial component of the embedding methods is the choice of a high-accuracy solver for the fragment problem. For practical molecular and materials applications, the dimensionality of the fragment is typically not very small; therefore to avoid the exponential scaling of full configuration interaction solvers,\cite{ASCI_Tubman_JCTC2020} CC theory is considered to be a viable, systematically improveable alternative with polynomial cost. It has been explored both in the context of DMFT \cite{Zhu_PRB19, Zhu_DMFT_PRX21}, its variants \cite{Shee2019} and DMET \cite{DMET_bootstrap_jcp_2016}. However, many of those approaches suffer from the lack of systematic accuracy and/or have insufficient theoretical robustness. 

A potentially well-defined alternative is to concentrate on coupling (high level) many-body CC with (lower level) perturbative methods in static embedding theory. Seminal efforts to combine CCSD (high level) and MP2 (low level) were the methods developed by Nooijen \cite{nooijen1999combining} and Sherrill \cite{bochevarov2005hybrid,bochevarov2006hybrid}, which demonstrated the importance of self-consistent coupling of the MP and CC amplitudes. In turn these methods relied on earlier work\cite{piecuch1993state,piecuch1994state} that  partitioned the cluster operator into internal and external components. Our recent work built on these and other developments\cite{MLCCJCP2014, MLCCjctc2021,cc_downfold_kowalski_JCP18,cc_downfold_kowalski23} to yield a self-consistent coupling between MP2 and CCSD (called MPCCSD) within a general MPCC wave function based static embedding theory\cite{mpcc_jcp24}. The MPCC approach is designed from the outset to enable well-defined, black-box partitioning of the system into fragment and environment amplitudes, based on an active orbital window. Alternatively spatial locality can additionally be used to consider fragments that are local to just an active region.

In this work, the same MPCC framework will be utilized to try to achieve the aforementioned goal of achieving quantitative accuracy for problems where CCSD(T) typically fails. In order to achieve this goal we have considered the iterative CCSDT method\cite{ccsdt3nogaJCP87} as a high-level fragment solver. The challenge is to design and assess one our more perturbation treatments that can describe the enviroment. The simplest option is to include only singles and doubles amplitudes in the environment, but one can also (and indeed we will show that one must also) include environment triples amplitudes.   



\section{Theory}

\subsection{Brief recap of the MPCC method} \label{Sec: theory_reacp}

In the MP-CC approach, we partition the total orbital space into a set of (active) fragment (F) orbitals and a set of (inactive) environment (E) orbitals. The fragment orbitals describe the chemically relevant region requiring higher accuracy, spanning only a few atoms, localized bonds, or a chemically relevant subspace of the full basis set.
The environment orbitals, on the other hand, will describe the remainder of the system.
A valid partition of a given orbital set $\{\phi_i\}_{i=1}^K$ is specified by disjoint index sets that assign each occupied and virtual orbital to either the fragment or the environment.
Assuming an $N$ electron system, the occupied space is,
\begin{equation}
O = O_{\rm E} \cup O_{\rm F},
\end{equation}
where $O=\{\phi_i:1\le i\le N\}$ and
\begin{equation}
O_{\rm E}=\{\phi_i:i\in\mathcal{I}_{\rm occ}^{\rm E}\}, \qquad
O_{\rm F}=\{\phi_i:i\in\mathcal{I}_{\rm occ}^{\rm F}\},
\end{equation}
with $\mathcal{I}_{\rm occ}^{\rm E}\cap\mathcal{I}_{\rm occ}^{\rm F}=\emptyset$ and
$\mathcal{I}_{\rm occ}^{\rm E}\cup\mathcal{I}_{\rm occ}^{\rm F}=[\![N]\!] = \{1,...,N\}$.
The virtual space is partitioned similarly:
\begin{equation}
V=V_{\rm E}\cup V_{\rm F}, \qquad
V=\{\phi_i:i>N\},
\end{equation}
where $V = \{ \phi_i : N < i < K \}$ and
\begin{equation}
V_{\rm E}=\{\phi_i:i\in\mathcal{I}_{\rm vir}^{\rm E}\}, \qquad
V_{\rm F}=\{\phi_i:i\in\mathcal{I}_{\rm vir}^{\rm F}\}
\end{equation}
with $\mathcal{I}_{\rm vir}^{\rm E}\cap\mathcal{I}_{\rm vir}^{\rm F}=\emptyset$ and $\mathcal{I}_{\rm vir}^{\rm E}\cup\mathcal{I}_{\rm vir}^{\rm F}=[\![K]\!] \setminus [\![N]\!].$
We denote the number of fragment and environment orbitals by $N_{\rm F}=|\mathcal{I}_{\rm occ}^{\rm F}|+|\mathcal{I}_{\rm vir}^{\rm F}|$ and $N_{\rm E}=|\mathcal{I}_{\rm occ}^{\rm E}|+|\mathcal{I}_{\rm vir}^{\rm E}|$, respectively. We slightly abuse the notation by using $O$ and $V$ to denote both the sets of occupied and virtual orbitals, respectively, and their cardinalities. The intended meaning should be clear from the context.

We will use the indices $i, j, k, l, ...$ for the orbitals in the occupied space ($O$) and $a, b, c, d, ...$ for the orbitals in the virtual space ($V$). For orbitals in the fragment occupied space ($O_F$) we use $I, J, K, L$, and for the fragment virtual space ($V_F$) we use $A, B, C, D$ indices.

We note that the choice of basis to construct the fragment space plays a vital role in obtaining the desired accuracy with this method, similar to the other embedding approaches. In our previous work \cite{mpcc_jcp24}, we have utilized the active valence active space \cite{AVASknizia2017} (AVAS) scheme to construct this space, as it provides an automatic approach. Within the AVAS scheme, we aim to isolate a set of orbitals of interest (such as d-orbitals of a transition metal) from the full set of orbitals of a specific problem. Typically, a minimal atomic orbital basis (MINAO) is chosen to represent the region of interest, as it closely aligns with the idea of chemically relevant bonding and antibonding orbitals in the valence region. On the other hand, for the full calculation, we can (and should) choose a larger atomic orbital basis set.
A special characteristic of the AVAS scheme is that higher angular momentum atomic orbitals can be trivially added to the active space by choosing a slightly larger basis set than the minimal one. This feature allows us to capture the strongest dynamical correlations within the active space and also provides a means to increase the active space size systematically.             

Once a valid partition is made, we employ an exponential wavefunction parametrization,
\begin{equation}
|\Psi\rangle = e^{T}|\Phi_0\rangle, \qquad 
T=\sum_{n\ge 1} T_n ,
\end{equation}
with
\begin{equation}
\begin{aligned}
T_n
&= \frac{1}{(n!)^2} \sum_{i_1\cdots i_n}\sum_{a_1\cdots a_n}
t_{i_1\cdots i_n}^{a_1\cdots a_n}\,
a_{a_1}^\dagger\cdots a_{a_n}^\dagger a_{i_n}\cdots a_{i_1}\\
&=\sum_{i_1\cdots i_n}\sum_{a_1\cdots a_n}
t_{i_1\cdots i_n}^{a_1\cdots a_n}\,
X_{i_1\cdots i_n}^{a_1\cdots a_n},
\end{aligned}
\end{equation}
where $X$ are particle-hole excitation operators \cite{Cizek_CC}. We define the fragment contribution at excitation rank $n$ as the restriction to fragment indices, i.e.,
\begin{equation}
T_n^{\rm F}=
\frac{1}{(n!)^2} \sum_{\substack{i_1,\ldots,i_n\in \mathcal I_{\rm occ}^{\rm F}\\
a_1,\ldots,a_n\in \mathcal I_{\rm vir}^{\rm F}}}
t_{i_1\cdots i_n}^{a_1\cdots a_n}\,
X_{i_1\cdots i_n}^{a_1\cdots a_n},
\end{equation}
and the environment contribution as its complement, i.e.,
\begin{equation}
T_n^{\rm E}=T_n-T_n^{\rm F}.
\end{equation}
Equivalently, $T^{\rm F}=\sum_{n\ge 1}T_n^{\rm F}$ and $T^{\rm E}=T-T^{\rm F}$. We denote by ${\bf t}^{\rm F}$ and ${\bf t}^{\rm E}$ the collections of amplitudes appearing in $T^{\rm F}$ and $T^{\rm E}$, respectively.
Note that $T^{\rm F}$ collects excitation operators that act {\it exclusively} within the fragment orbital subspace. In contrast, $T^{\rm E}$ comprises all remaining amplitudes, i.e., (i) ``purely'' environmental excitations involving only environment orbitals and (ii) ``mixed'' fragment-environment excitations. The mixed terms capture correlations that couple fragment and environment degrees of freedom, e.g., excitations from fragment occupied orbitals into environment virtual orbitals, or simultaneous excitations involving occupied orbitals from both regions.


In the spirit of quantum embedding, we will compute the cluster operators $T^{\rm F}$ and $T^{\rm E}$ at different levels of accuracy. Specifically, we determine the fragment cluster operator $T^{\rm F}$ using the standard CC formalism (the high-level, HL, theory in quantum embedding parlance), while treating the environment cluster operator $T^{\rm E}$ at a perturbative level (the lower-level, LL, theory).
Put differently, the similarity-transformed Hamiltonian that enters the CC projection equations is treated differently in the fragment and environment theories. For the fragment, we use the full similarity-transformed Hamiltonian, i.e.,
\begin{equation}
\overline{H}({\bf t}^{\rm F}; {\bf t}^{\rm E})
= e^{-T^{\rm E}-T^{\rm F}} H e^{T^{\rm F}+T^{\rm E}},
\end{equation}
whereas for the environment, we employ a lower-order perturbative approximation, denoted $\widetilde{H}({\bf t}^{\rm E}; {\bf t}^{\rm F})$. The latter is theoretically more involved; a detailed discussion of the low-level Hamiltonians proposed and investigated here follows below.
In overview, the MP-CC approach leads to two sets of coupled projective equations
\begin{align}
     \langle \Phi_\mu^{\rm F}| \overline{H}^{\rm F}({\bf t}^{\rm F}; {\bf t}^{\rm E})| \Phi_0 \rangle = {}& 0, \label{Eq:projF} \\
     \langle \Phi_\mu^{\rm E}| \widetilde{H}^{\rm E}({\bf t}^{\rm E}; {\bf t}^{\rm F}) | \Phi_0 \rangle = {}& 0, \label{Eq:projEB}
\end{align}
where $|\Phi_\mu^{\rm F}\rangle$ and $|\Phi_\mu^{\rm E}\rangle$ denote excited Slater determinants in the fragment and environment space, respectively.
These coupled residual equations can be obtained from the stationarity of the MP-CC Lagrangian,
\begin{equation}
\label{Eq:LagrangianGen}
\begin{aligned}
\mathcal{L}({\bf t},\boldsymbol{\lambda}) 
&= \langle \Phi_0 | \overline{H} | \Phi_0 \rangle + \langle \Phi_0 | \Lambda^{\rm F} \overline{H}^{\rm F}({\bf t}^{\rm F}; {\bf t}^{\rm E}) | \Phi_0 \rangle\\
&\quad + \langle \Phi_0 | \Lambda^{\rm E} \widetilde{H}^{\rm E}({\bf t}^{\rm E}; {\bf t}^{\rm F}) | \Phi_0 \rangle, 
\end{aligned}    
\end{equation}
where $\Lambda^{Y}=\sum_{\mu\in Y}\lambda_\mu^{Y}X_\mu^{Y}$ for $Y\in\{{\rm F},{\rm E}\}$, and the first term in the Lagrangian is the coupled cluster energy for the total system (fragment + environment); therefore the $\overline{H}$ remained unlabeled. 

We emphasize that both $\overline{H}^{\rm F}$ and $\widetilde{H}^{\rm E}$ in Eqs.~\eqref{Eq:projF} and~\eqref{Eq:projEB} depend on the full set of cluster amplitudes, i.e., ${\bf t}_{\rm F}$ {\it and} ${\bf t}_{\rm E}$. However, when solving the fragment residual equations, ${\bf t}_{\rm E}$ is held fixed and ${\bf t}_{\rm F}$ is updated, whereas in solving the environment residual equations, ${\bf t}_{\rm F}$ is held fixed and ${\bf t}_{\rm E}$ is updated. Consequently, the two sets of equations are coupled and can advantageously be self-consistently solved using a double-loop procedure consisting of \textit{macro}- and \textit{micro}-iterations.
In the \textit{macro}-iterative step, we first solve the low-level amplitude equations to obtain the environment cluster amplitudes, $T^{\rm E}$. We then construct the screened interaction for the fragment problem,
\begin{equation}
\label{eq:WF}
    W^{\rm F} = e^{-T^{\rm E}} H e^{T^{\rm E}},
\end{equation}
by similarity-transforming the bare Hamiltonian with $T^{\rm E}$ and subsequently restricting all indices to the fragment subspace. This operation typically generates effective Hamiltonian matrix elements with higher particle-rank (many-body) contributions, i.e., although $H$ is a two-body operator, $W^{\rm F}$ generally contains effective three-body and higher-body terms.
In the \textit{micro}-iterative step, we solve the fragment amplitude equations using $W^{\rm F}$. The resulting fragment amplitudes, ${\bf t}_{\rm F}$, enter the low-level amplitude equations and thereby update the next \textit{macro}-iterative step. We also note that the $W^F$ is typically denoted as a downfolded Hamiltonian, and Eq.~\eqref{eq:WF} provides a static approach to build this quantity. A downfolded Hamiltonian can encapsulate the dynamical correlation effects of the environment into the fragment.  As the number of degrees of freedom of the fragment problem is much smaller than the total system, we can treat the downfolded Hamiltonian with a higher-accuracy solver. A similar static approach was taken previously by Kowalski \textit{et al.} to construct both unitary \cite{cc_downfold_karol19, DUCC_Kowalski_JCTC2020} and non-unitary \cite{cc_downfold_kowalski_JCP18, cc_downfold_kowalski23} downfolded Hamiltonians, and Evangelista \textit{et al.} \cite{DSRG_downfolding_PRX2023} used a different framework, namely the driven similarity renormalization group (DSRG) approach, for this construction. However, unlike our work, these studies solved the downfolded Hamiltonian in isolation, that is, not self-consistently within an embedding framework.

In this work, we primarily focus on obtaining a higher accuracy solver for the fragment. Therefore, the necessary modifications to the solver will be elaborated in the following sections. We have also found that certain improvements to the low-level solver are necessary to achieve the desired accuracy from a high-level solver. Those improvements will also be discussed in the following sections. 

\subsection{Low-level methods for environment singles and doubles} \label{Sec:theory_lowlevel}

The lower-order expansion in $\widetilde{H}^{\rm E}$ allows us to choose various options for gaining a computational advantage, as was elaborated in our earlier work \cite{mpcc_jcp24}. In the following, we will briefly explain the most useful approximation made, namely the relaxed scheme, which led to obtaining both qualitative and quantitative accuracy. Orbital relaxation of the environment was identified as vital,\cite{nooijen1999combining,bochevarov2005hybrid,bochevarov2006hybrid,mpcc_jcp24}, and according to the Thouless theorem \cite{thouless1960stability}, the $e^{T_1}$ part of the CC ansatz can capture that effect. Therefore, a perturbation theory was defined in terms of the $e^{T_1}$ transformed Hamiltonian, where the following partitioning was used:
\begin{equation}
    \widetilde{H} = e^{-T_1} H e^{T_1}
    = E_{\mathrm{cl}}\,I + \tilde{F} + \tilde{V} \label{Eq:mp_partition},
\end{equation}
where $E_{\mathrm{cl}}$ is the scalar (zero-body) contribution. Following an MP-type partitioning, we take $\tilde{F}$ as the zeroth-order part and $\tilde{V}$ as the first-order (fluctuation) contribution.
We then define the first-order environment equations for the singles and doubles amplitudes as
\begin{footnotesize}
\begin{align}
    \langle \Phi_\mu^{\rm E} | \tilde{F} + [\tilde{F}, T_2] + [\tilde{F}, T_1] | \Phi_0 \rangle &= 0,\,   
    \mu = {\binom{a}{i}} \setminus {\binom{A}{I}}
     ,\\
    \langle \Phi_\mu^{\rm E} | \tilde{V} + [\tilde{F}, T_2] | \Phi_0 \rangle  &= 0,\,\mu = {\binom{ab}{ij}} \setminus {\binom{AB}{IJ}}.
\end{align}
\end{footnotesize}
We refer to the embedding scheme that employs this set of environment equations as the MP1CC method.

In this work, we are also going to employ an improved variant of the environment treatment by including second-order terms in the perturbative equations, i.e.,
\begin{footnotesize}
\begin{align}
    \langle \Phi_\mu^{\rm E} | \tilde{F} + [\tilde{F}, T_1] + [\tilde{F}+ \tilde{V}, T_2] | \Phi_0 \rangle &= 0,~ \mu = {\binom{a}{i}} \setminus {\binom{A}{I}} , \\
    \langle \Phi_\mu^{\rm E} | \tilde{V} + [\tilde{F}, T_2] + [\tilde{V}, T_2] | \Phi_0 \rangle &= 0,~ \mu = {\binom{ab}{ij}} \setminus {\binom{AB}{IJ}}.
\end{align}
\end{footnotesize}
We refer to the embedding scheme that employs this second-order set of environment equations as the MP2CC method. This approach increases the cost of the low-level step to $\mathcal{O}(N^6)$, compared to $\mathcal{O}(N^5)$ for MP1CC. As we shall see, this higher cost also yields improved accuracy in some strongly correlated problems.

\subsection{CCSDT as a fragment solver} \label{Sec:theory_fragsolver}

In this work, our goal is to solve the fragment problem using a method beyond CCSD. The fragment is described by the downfolded Hamiltonian mentioned in  Eq.~\eqref{eq:WF}, where $T^{\rm E}$ consists of $T^{\rm E}_1$, $T^{\rm E}_{2}$ amplitudes discussed above, defining:
\begin{equation}
    W^{\rm F} = e^{-T_1^{\rm E} -T_2^{\rm E}} H e^{T_1^{\rm E} + T_2^{\rm E}}. \label{Eq:downfold_t2e}
\end{equation}
The wavefunction ansatz for the fragment problem is further augmented with the triples amplitudes, such that it is CCSDT:
\begin{equation}
    |\Psi^{\rm F}\rangle = e^{T^{\rm F}}|\Phi_0\rangle
    = e^{T^{\rm F}_1+T^{\rm F}_2+T^{\rm F}_3}|\Phi_0\rangle,
\end{equation}
where $T^{\rm F}_3$ denotes the fragment triples operator. This anstaz then substituted into Eq.~\eqref{Eq:projF} to derive the relevant projection equations. We note that the fragment Hamiltonian, Eq.~\eqref{Eq:downfold_t2e}, formally contains up to three-body terms. However, by suitable reordering of the tensor operations, they do not explicitly appear.


\subsection{Complete neglect of environment triples}\label{sec: MPCCSDT_zero}

The environment triples amplitudes are treated perturbatively, in analogy with the truncated singles-doubles environment method. The definition of $\widetilde{H}^{\rm E}$ is not straightforward when triple excitations are included, and we therefore investigate different approaches. In the simplest possible approach, we follow an argument from our previous work~\cite{mpcc_jcp24}: the MP1-level perturbation theory used to determine the environment amplitudes does not generate any $T_3$ contributions for a Hamiltonian that is at most two-body. We therefore set $T^{\rm E}_3 = 0$.
In the CCSDt method by Adamowicz \textit{et.~al} \cite{adamowicz1998state}, a similar choice was made, and only active space triples were considered to be nonzero within CCSDT. We will therefore refer to this MPCC approach as MPCCSDt. 
The above method has an iterative $\mathcal{O}(V_{F}^5O_F^3)$ cost for the fragment solver.

However, as noted previously in the context of CCSDt~\cite{ShenLi_CCSDt_JCP2010,CCt3_JCTCShen2012} and corroborated by our numerical results here, some treatment of environment triples ($T^{\rm E}_3$) is essential to obtain chemically accurate energetics. To that end, we have designed and implemented two perturbative schemes to describe the $T^{\rm E}_3$ amplitudes without fully evaluating their effect on the fragment Hamiltonian. One is single-shot non-iterative, denoted as MPCCSDT(pt), where the environment triples are fully decoupled from the fragment amplitude equations. The other allows leading order self-consistent coupling between the envionment triples and fragment amplitudes based on an iterative perturbative procedure, which we denote as MPCCSDT(it). They are discussed in the following subsections.

\subsection{Single-shot perturbation theory for environment triples}\label{sec: MPCCSDT_pt}

For the evaluation of environment triples amplitudes, we used a perturbative and noniterative formulation, as the lowest-order construction of these amplitudes requires an $\mathcal{O}(N^7)$ computational cost. In the interest of constructing a noniterative algorithm, we define $W_{\rm F}$ to be independent of $T_{3}^{\rm E}$, so that the equations used to evaluate the fragment amplitudes are decoupled from the environment triple amplitudes. To derive the equation for $T_{3}^{\rm E}$, we first apply the same similarity transformation of the Hamiltonian used for the singles–doubles amplitude equations, namely by $e^{T_1}$, and then partition the Hamiltonian in exactly the same manner as in Eq.~\eqref{Eq:mp_partition}.

Furthermore, unlike the SD equations, we exclude the $\tilde{F}_{ov}$ term from the zeroth-order Hamiltonian and consider it as a first-order term, as we want to evaluate $T_{\rm 3}^{\rm E}$ in a non-iterative manner. Otherwise, the amplitude equation for $T_2^{\rm E}$ would contain an additional commutator term, $\left[\tilde{F}_{ov}, T_3^{\rm E}\right]$, which would couple the $T_3^{\rm E}$ and $T_2^{\rm E}$ amplitude equations. Thus, the zeroth-order Hamiltonian consists of only $\tilde{F}_{oo}$ and $\tilde{F}_{vv}$ terms. The lowest-order environment triples amplitudes are then obtained from the following projection equation:
\begin{footnotesize}
\begin{align}
 \langle \Phi_\mu ^{E} |
 W_E + [\tilde{F}_{oo} + \tilde{F}_{vv}, T_3^{(1)}] |\Phi_0\rangle &= 0;\, \mu = {\binom{abc}{ijk}} \setminus {\binom{ABC}{IJK}}
 \label{Eq:T3_ampl},\\
  \left[ \tilde{V}, T_2 \right] + \left[ \tilde{V}, T_3 \right] &= W_E
 \label{Eq: constW}
\end{align}
\end{footnotesize}
In this work, we further neglected the second commutator term containing the $T_{\rm 3}$ amplitudes, so that it does not generate a set of coupled equations, which is difficult to solve in a non-iterative manner. We also assume that its contribution is smaller than that of the commutator term containing the $T_{\rm 2}$ amplitudes. It should be noted that the terms in Eq.~\eqref{Eq: constW} are second-order perturbative terms. Thus, Eq.~\eqref{Eq:T3_ampl} is not fully consistent with respect to the perturbation order when first-order amplitude equations are used for the SD ones. 
However, it is consistent when second-order amplitude equations are considered in that case. Nevertheless, we compare both combinations numerically, as they entail different computational costs for the SD equations.

Now, the contribution of $T_{\rm 3}^{\rm E}$ to the total energy can be derived as follows. We first notice that the contribution of $T_{\rm 3}^{\rm E}$ amplitudes to the residuals of $T_1$ and $T_2$ amplitudes was not taken into account, which for a full CCSDT computation would have vanished. This non-zero residual will contribute to the total energy, and it can be extracted from the CC Lagrangian:
\begin{equation}
\begin{aligned}
    \mathcal{L}_{\rm E} ={}& \langle \Phi_0 | (1 + \Lambda_1 + \Lambda_2) [\tilde{V}, T_{3}^{\rm E}] | \Phi_0 \rangle  \\
    ={}& \langle \Phi_0 | [(\Lambda_2 \tilde{V})_{c}, T_{\rm 3 }^{\rm E}] + [(\Lambda_1 \tilde{V})_{dc}, T_{\rm 3}^{\rm E}] \\
    &\quad + [(\Lambda_2 \tilde{F})_{dc}, T_{3}^{\rm E}]  | \Phi_0 \rangle \\
    ={}& (\tilde{V}^{ek}_{bc} \lambda^{ij}_{ae} - \tilde{V}^{ij}_{am} \lambda^{mk}_{bc} + \tilde{V}^{ij}_{ab} \lambda^k_c + \tilde{F}^i_a \lambda^{jk}_{bc})_E T^{abc, \rm E}_{ijk}  \\
    ={}& Y^{ijk}_{abc, \rm E} T^{abc, \rm E}_{ijk} \label{Eq:pert_t}
\end{aligned} 
\end{equation}
In the above equations, $(\cdots)_c$ denotes the connected terms and $(\cdots)_{dc}$ denotes the disconnected terms.

We further approximated $\Lambda_1 = T_1^\dagger$ and $\Lambda_2 = T_2^\dagger$ as leading-order contributions to the CC left eigenvectors. We will henceforth call this model MPCCSDT, that is, where all triples amplitudes are represented.    


We can furthermore enhance our Lagrangian with this perturbative energy correction term and the amplitude equation Eq.~\eqref{Eq:T3_ampl} for the environment triples amplitudes, which constitutes the full Lagrangian for the MPCCSDT theory.

The energy expression derived in Eq.~\eqref{Eq:pert_t} has structural similarities with the $\Lambda$CCSD(T) method by Kucharski and Bartlett \cite{LCCSD(T)_BartlettJCP08} and the triples approximated variant of the CCSD(2) method, namely CCSD(2)$_T$ by Gwaltney and Head-Gordon \cite{CCSD(2)GwaltneyMHGJCP01}. However, those formulations differ from each other, and from the method discussed here in two ways: a) what form of the Hamiltonian, that is, similarity transformed or not, enters the equation, and b) what definition is used for $T_3$. More specifically, in Eq.~\eqref{Eq:T3_ampl} we evaluate only environment amplitudes, where a fixed set of fragment $T_3$ amplitudes contributes a regularization term as will be shown in Eq.~\eqref{Eq:T3_ampl_LL}. Another distinction is that the $T_1$ and $T_2$ amplitudes that appear in Eq.~\eqref{Eq:pert_t} and \eqref{Eq:T3_ampl} are optimized in the presence of $T_3^F$ amplitudes. In their method of moment coupled cluster (MMCC) \cite{MMCC_Piecuch01102002} formalism by Piecuch \textit{et. al.} evaluate a similar energy correction due to the $T_3$ amplitudes, where an orbital non-invariant formula is derived based on the Epstein-Nesbet type denominator. Later, the same formalism was extended to evaluate energy correction on top of the CCSDt formalism, and the resulting method is called CC(t;3)\cite{CCt3_JCTCShen2012}. Perhaps, CC(t;3) is the most closely related theory as our environment triples. Though some of the distinctions already mentioned above hold for the CC(t;3) method as well. Moreover, in CC(t;3) formalism Hamiltonian moment was used for the $e^{T_1 + T_2}$ transformed Hamiltonian. In comparison, we used only a $e^{T_1}$ transformed Hamiltonian to evaluate the $T_3$ amplitudes in Eq.~\eqref{Eq:T3_ampl}.

We will now elaborate on the solution strategy of Eq.~\eqref{Eq:T3_ampl}: for a fixed set of $T_{3}^{\rm F}$ amplitudes, we aim to find the $T_{3}^{\rm E}$ amplitudes. Since $\tilde{F}$ is generally a non-diagonal matrix, Eq.~\eqref{Eq:T3_ampl} presents a system of coupled equations of different $T_3$ amplitudes. Therefore, an iterative approach will be more suitable to solve this system of equations. However, this will incur an iterative $\mathcal{O}(N^7)$ cost of the above equations. In order to eliminate the iterative cost, we cast the equation in the following manner:
\begin{equation}
\begin{aligned}
    &\langle \chi_{ijk, E}^{abc} |W_{ijk}^{abc} + [\tilde{F}, T_{3}^{LL}] + [\tilde{F}, T_{3}^{\rm F,HL} - T_{3}^{\rm F, LL}] |\Phi_0\rangle \\
    &~= \langle \chi_{ijk, E}^{abc} |\hat{W}_{ijk}^{abc} + [\tilde{F}, T_{3}^{LL}] |\Phi_0\rangle\\
    &~= 0 \label{Eq:T3_ampl_modified}   
\end{aligned}
\end{equation}
where, LL stands for the low-level method and HL stands for the high-level method or the previously obtained $T_{\rm 3}^{\rm F}$ amplitudes,  $T_3 = T_{\rm 3}^{\rm F} + T_{\rm 3}^{\rm E}$ and $T_{3}^{\rm F, LL}$ amplitudes are obtained from the following equations:
\begin{equation}
 \langle \chi_{ijk}^{abc} |W_{ijk}^{abc} + [\tilde{F}, T_{3}^{LL}] |\Phi_0\rangle = 0 \label{Eq:T3_ampl_LL},
\end{equation}

Now, Eq.~\eqref{Eq:T3_ampl_LL} can be solved by transforming the $\tilde{F}$ matrix into the semicanonical basis, and Eq.~\eqref{Eq:T3_ampl_modified} can also be solved on the same semicanonical basis when we satisfy 
\begin{equation}
    \langle \chi_{ijk, F}^{abc} |W_{ijk}^{abc} + [\tilde{F}, T_{3}^{LL}] |\Phi_0\rangle = 0 \label{Eq:T3_ampl_fragment_LL}
\end{equation}
in the original basis of that equation. This will modify Eq.~\eqref{Eq:T3_ampl_modified} in the following way
\begin{equation}
    \langle \chi_{ijk}^{abc} |\hat{W}_{ijk}^{abc} + [\tilde{F}, T_{3}^{LL}] |\Phi_0\rangle = 0 \label{Eq:T3_ampl_final}, 
\end{equation}
that is, we do not have any index restrictions now. As a result, Eq.~\eqref{Eq:T3_ampl_final} can be solved in a non-iterative manner by transforming it to a semicanonical basis, similar to Eq.~\eqref{Eq:T3_ampl_LL}. Thus, the evaluation of $T_{\rm 3}^{\rm E}$ can be summarized in four steps:

\begin{enumerate}
    \item build $W_{ijk}^{abc}$ from Eq.~\eqref{Eq: constW} using ERIs and converged $T_1$, $T_2$ amplitudes,  
    \item solve Eq.~\eqref{Eq:T3_ampl_LL} in a semicanonical basis to obtain $T_{3}^{\rm F, LL}$ amplitudes,
    \item construct $\hat{W}_{ijk}^{abc}$ by using $T_{3}^{\rm F, LL}$ and $T_{3}^{\rm F, HL}$ amplitudes: $\hat{W}_{ijk}^{abc} = W_{ijk}^{abc} + ([\tilde{F}, T_{3}^{\rm F,HL}] - [\tilde{F}, T_{3}^{\rm F, LL}])$,
    \item solve Eq.~\eqref{Eq:T3_ampl_final} in a semicanonical basis to obtain $T_{\rm 3}^{\rm E}$ amplitudes.
\end{enumerate}

None of the above steps involves an iterative cost. An important aspect of this non-iterative method is that the semi-canonical basis obtained by diagonalizing the $\tilde{F}$ matrix is biorthogonal, as $\tilde{F}$ is not hermitian.  

Since the above scheme takes all triples (fragment and environment) amplitudes into consideration and the $T_{3}^{\rm E}$ amplitudes are treated perturbatively, we will denote this method as MPCCSDT(pt).

\subsection{Iterative perturbation theory for environment triples}

In the perturbative treatment of environment triples amplitudes, we have neglected the influence of inactive triples on the fragment singles, doubles, and triples amplitudes. We consider that this particular kind of coupling may have significant contributions in the strongly correlated cases. To account for this coupling, we first solve the perturbative amplitude Eq.~\eqref{Eq:T3_ampl} for $T_{3} ^{\rm E}$ amplitudes. The solution strategy remains the same as we have outlined in the previous section. We furthermore construct a modified fragment Hamiltonian by applying an additional similarity transformation of $e^{T_{3}^{\rm E}}$ operator to the Hamiltonian. Therefore, the modified definition of the downfolded fragment Hamiltonian is:
\begin{equation}
    W_F = e^{-T_{1}^{\rm E} -T_{2}^{\rm E} - T_{3}^{\rm E}} H e^{T_{1}^{\rm E} + T_{2}^{\rm E} + T_{3}^{\rm E}} \label{Eq: downfold_T3E}
\end{equation}

We apply this transformation in every \textit{macro} iterative step to update the fragment Hamiltonian, $W_F$. Therefore, we will refer to this method as MPCCSDT(it). A closer analysis of Eq.~\eqref{Eq:pert_t} reveals that the energy contribution due to environment triples terms will be negligible now, which we will explain below. The amplitude equations for the $T_{1}^{\rm F}$, $T_{2}^{\rm F}$ and $T_{3}^{\rm F}$ cluster operators in the MPCCSDt and MPCCSDT(pt) methods are formulated using the downfolded Hamiltonian defined in Eq.~\eqref{Eq:downfold_t2e}. Upon augmenting the additional terms to the downfolded Hamiltonian via Eq.~\eqref{Eq: downfold_T3E}, the fragment amplitude equations acquire extra residual contributions, that is, $\tilde{R} = R + R'$, where $R'$ is the additional contribution to the previous residual R. By construction, $\tilde{R}$ corresponds to the fragment subset of terms from the full CCSDT amplitude equation. In the self-consistent MPCCSDT(it) formulation, the fragment amplitudes are obtained by solving amplitude equations iteratively until $\tilde{R} \rightarrow 0$, as opposed to $R \rightarrow 0$ in the MPCCSDT(pt) and MPCCSDt methods. The term $R'$ represents the fragment subset of terms of the $[ \tilde{V}, T_3^{\rm E} ]$ commutator appearing in Eq.~\eqref{Eq:pert_t}. This contribution vanishes for the MPCCSDT(it) method by virtue of the self-consistency condition mentioned above. However, in this method, we still neglect the total energy contribution from the environment subset of terms of the $[ \tilde{V}, T_3^{\rm E} ]$ commutator. Thus, the influence of $T_{3}^{\rm E}$ amplitudes on the $T_{2}^{\rm E}$ and $T_{1}^{\rm E}$ amplitudes was not captured, which we assume to be negligible and is supported by numerical tests as well. We note that an analogous approximation is employed in the MPCCSD method.   

The computational cost of evaluating the influence of $T_{\rm 3}^{\rm E}$ amplitudes on fragment amplitudes remains relatively low. Notably, the most expensive term in the CCSDT method, the particle-particle ladder (PPL) term, can now be evaluated with a cost of $V_F^3 O_F^3 V^2$, compared to the previous $V^5 O^3$. It also means that we need to evaluate only a subset of $T_{\rm 3}^{\rm E}$ amplitudes, for which the computational scaling will be $\mathcal{O}(V^3V_F O_F^2O)$. Additionally, the construction of these terms is required only during the macro iterative step. Consequently, we have a prefactor of the number of macro iterations to the above computational scaling. However, the number of macro iterations is typically quite small, ranging from 5 to 10.   

We have summarized the computational cost and the memory requirements of each of the methods in Table \ref{tab:compute-memory}.

\begin{table*}[t]
\centering
\begin{tabular}{lcccccc}
\hline
\textbf{Method} & \multicolumn{2}{c}{\textbf{Macro}} & \multicolumn{2}{c}{\textbf{Micro}} & \multicolumn{2}{c}{\textbf{Perturbative}} \\
\hline
 & Compute & Memory & Compute & Memory & Compute & Memory \\

UMP1CCSDt       & $\mathcal{O}(O^2V^2X)$ & $\max$($\mathcal{O} (N^2 X)$, $\mathcal{O}(O^2 V^2)$)  & $\mathcal{O}(V_F^5O_F^3)$ & $\mathcal{O} (V_F^3 O_F^3)$ & None & None   \\
UMP1CCSDT(pt)   & $\mathcal{O}(V^2 O^2 X)$ & $\mathcal{O}(V^3 O)$  & $\mathcal{O}(V_F^5O_F^3)$  & $\mathcal{O} (V_F^3 O_F^3)$ & $\mathcal{O} (V^4O^3)$ & $\max (\mathcal{O}(V^3 O), \mathcal{O} (V_F^3 O_F^3))$ \\
UMP1CCSDT(it)   & $\mathcal{O}(V^3V_F O_F^2O)$ & $\mathcal{O}(V^2 O V_F O_F^2 )$ & $\mathcal{O}(V_F^5O_F^3)$ & $\mathcal{O} (V_F^3 O_F^3)$ & None & None \\
UMP2CCSDt        & $\mathcal{O}(O^2V^4)$ & $\mathcal{O}(V^4)$  & $\mathcal{O}(V_F^5O_F^3)$ & $\mathcal{O} (V_F^3 O_F^3)$ & None & None \\
UMP2CCSDT(pt)   & $\mathcal{O}(O^2V^4)$  & $\mathcal{O}(V^4)$  & $\mathcal{O}(V_F^5O_F^3)$ & $\mathcal{O} (V_F^3 O_F^3)$ & $\mathcal{O}(V^4 O^3)$  & $\max (\mathcal{O}(V^3 O), \mathcal{O} (V_F^3 O_F^3))$  \\
UMP2CCSDT(it)   & $\mathcal{O}(V^3V_F O_F^2O)$ & $\mathcal{O}(V^2 O V_F O_F^2 )$ & $\mathcal{O}(V_F^5O_F^3)$ & $\mathcal{O} (V_F^3 O_F^3)$ & None & None \\
\hline
\end{tabular}
\caption{Compute and memory scaling for different methods at different steps of the methods. X denotes the size of the auxiliary basis set. We will denote the dimensionality of various sets of orbitals by the set itself, such as, $O_F \equiv \lvert O_F \rvert$, $V_F \equiv \lvert V_F \rvert$, etc.}
\label{tab:compute-memory}
\end{table*}



\section{Implementation}

All the methodological developments described in the previous sections have been implemented in the \texttt{PySCF}~\cite{sun2015libcint,sun2018pyscf,sun2020recent} quantum chemistry package. The implementation of these variants is substantially more memory intensive and computationally more demanding than the MPCCSD method. We therefore employ three-center density-fitted (DF) electron repulsion integrals (ERIs) instead of the four-center ones when evaluating the triples amplitudes. In addition, we store only the fragment $T_3$ amplitudes in memory, while the environment $T_3$ amplitudes are handled using a disk-driven approach.
The implementation of the MPCCSDT(it) scheme also requires the construction of fragment Hamiltonian matrix elements that involve tensor contractions between the bare Hamiltonian and the environment triples amplitudes ($T_{3}^{\rm E}$). These terms are likewise implemented using a disk-driven approach. To implement Eq.~\eqref{Eq:T3_ampl}, which is the low-level amplitude equation used to evaluate the $T_{3}^{\rm E}$ amplitudes, we introduce an approximation. As discussed in Sec. \ref{sec: MPCCSDT_pt}, a semicanonical basis is used to solve this equation perturbatively. However, this semicanonical basis is biorthogonal in nature because the Hamiltonian entering this equation is non-Hermitian.
We do not perform an exact implementation using two different sets (left and right)
of semicanonical orbitals for the transformation of the matrix elements, as has been done in the CCSD(2) work by Gwaltney \textit{et. al.} \cite{CCSD(2)_Gwaltney_MHG2000}. Instead, we symmetrize the $\tilde{F}$ matrix and construct the semicanonical basis from its diagonalization. This approach also avoids the appearance of complex semicanonical orbitals and complex eigenvalues and denominators. We have tested the validity of this approximation by comparing it with an iterative implementation of the same equation, which does not require the construction of a semicanonical basis. We observe a deviation of approximately $10^{-5}$~H, which is negligible. In future work, we aim to obtain a Hermitian Hamiltonian for the environment problem to avoid all the problems mentioned here.


\section{Results}
We will analyze various aspects of the newly proposed methods with two sets of applications: a) potential energy curves of the N$_2$ and F$_2$ molecules to examine the error versus full CCSDT at different bond distances; b) bond dissociation energies (BDE), and total atomization energies (TAE) of transition metal hydrides (TMH) and a selected set of molecules from the W4-11 dataset \cite{W4_KARTON_CPL2011}, respectively.

\subsection{Potential energy curves:}
We will assess our triples embedding models for N$_2$ and F$_2$ potential energy curves by measuring deviations against unrestricted CCSDT (UCCSDT). For both N$_2$ and F$_2$, the UCCSD error is quite large, which allows us to meaningfully investigate the role of triples. The cc-pCVTZ basis is chosen to be the large target basis, so our reference calculations are UCCSDT/cc-pCVTZ. Following our previous work, we have chosen the AVAS scheme to construct the fragment basis. Within this scheme, we chose cc-pVDZ to define the space of active orbitals,
which previously yielded good performance in the MPCCSD method.

\begin{figure}
    \centering
    \includegraphics[width=0.5\textwidth]{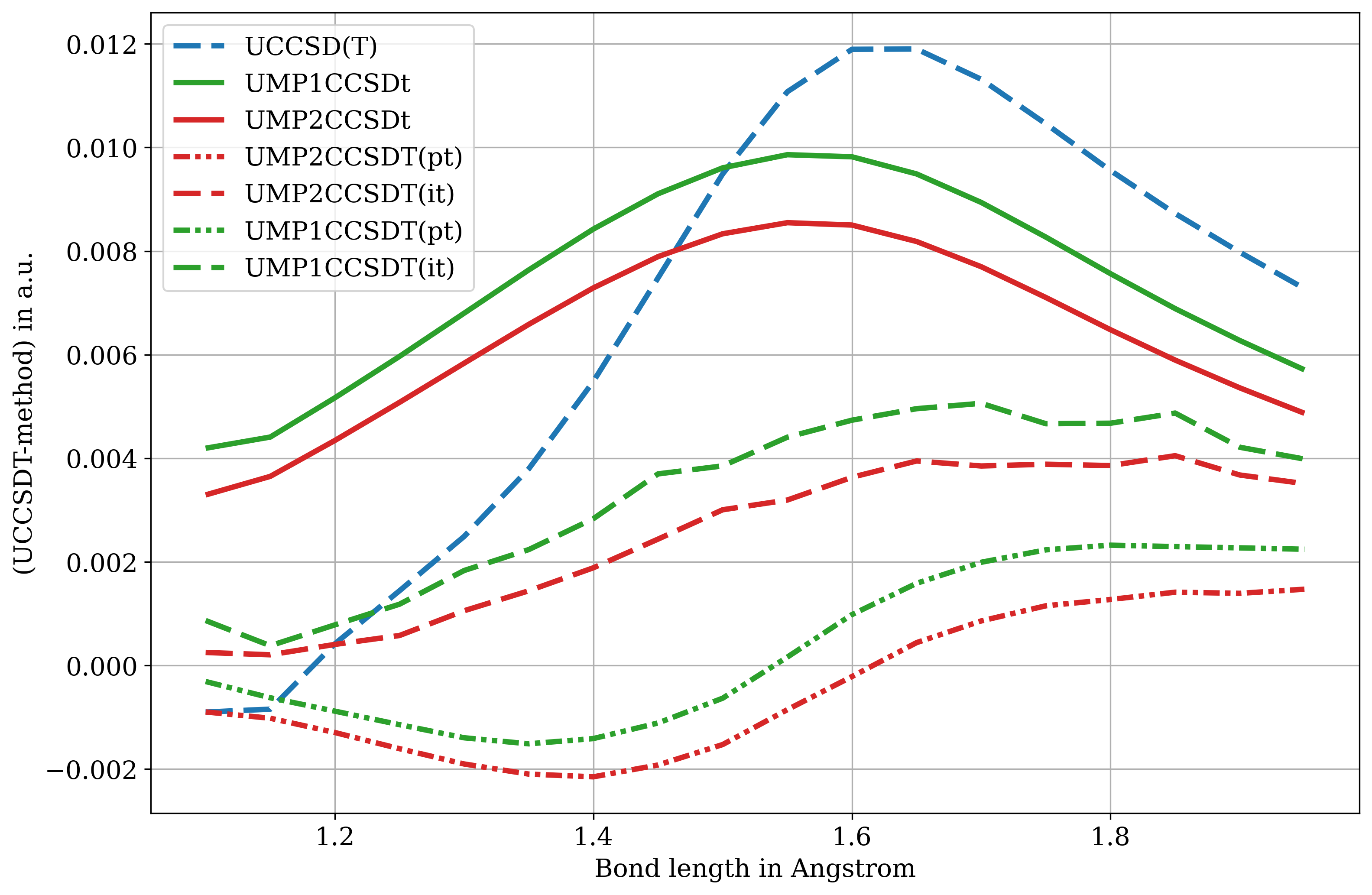}
    \caption{Electronic energy error with respect to UCCSDT along a PEC for N$_2$ in the cc-pCVTZ basis set. AVAS active space is constructed with cc-pVDZ basis set. We set the energy at dissociation as the zero energy for each method.}
    \label{fig:N2_pes}
\end{figure}

Fig. \ref{fig:N2_pes} plots the error in N$_2$ interaction energies ($E_{mol} - 2*E_{atom}$) for each method versus bond length for this triple bond stretching problem, using UHF orbitals to ensure correct dissociation. In the equilibrium region, the CCSD(T) method performs very well, with $< 1$ mH error. As we further stretch the bond, entering the spin-recoupling region, CCSD(T) exhibits poor behavior, and attains a non-parallelity error (NPE) of 13 mH. We also observe a first derivative discontinuity in the CCSD(T) error curve near the equilibrium region, at the Coulson-Fischer point, as expected for a perturbative method \cite{kurlancheek2009violations}.

Moving on to the embedding schemes, we observe that the UCCSD(T) error in the spin-recoupling region is significantly reduced for all the variants. With the MPCCSDt methods (solid curves), where we employ CCSDT only for the fragment, NPE is reduced by $\sim$ 2 mH (UMP1CCSDt) and $\sim$ 3.5 mH (UMP2CCSDt), which is encouraging. However, the remaining NPE is still chemically relevant, and is attributable to neglect of environment triples amplitudes. This is proven by including the environmental triples contribution perturbatively via the UMPCCSDT(pt) method (dot-dashed curves), which greatly reduces the error at every bond distance. Even then, the performance of our environmental triples scheme is not completely satisfactory, as a change in sign occurs in the error curve and a $\sim$ 3.5 mH NPE remains in the interaction energy curve. When we introduce the coupling between the fragment and the environment triples ($T_{3}^{\rm E}$) amplitudes via the UMPCCSDT(it) methods, the error curve looks better behaved, as there is no change in sign. Note that the UMPCCSDT(it) curves are not completely smooth, which we have found is due to using density fitted integrals in the evaluation of $T_{3}^{\rm E}$ amplitudes (Fig. \ref{fig:N2_pes_DF}). Note that the choice of the low-level method does not significantly affect the error curve, in comparison to other factors. 

\begin{figure}
    \centering
    \includegraphics[width=0.5\textwidth]{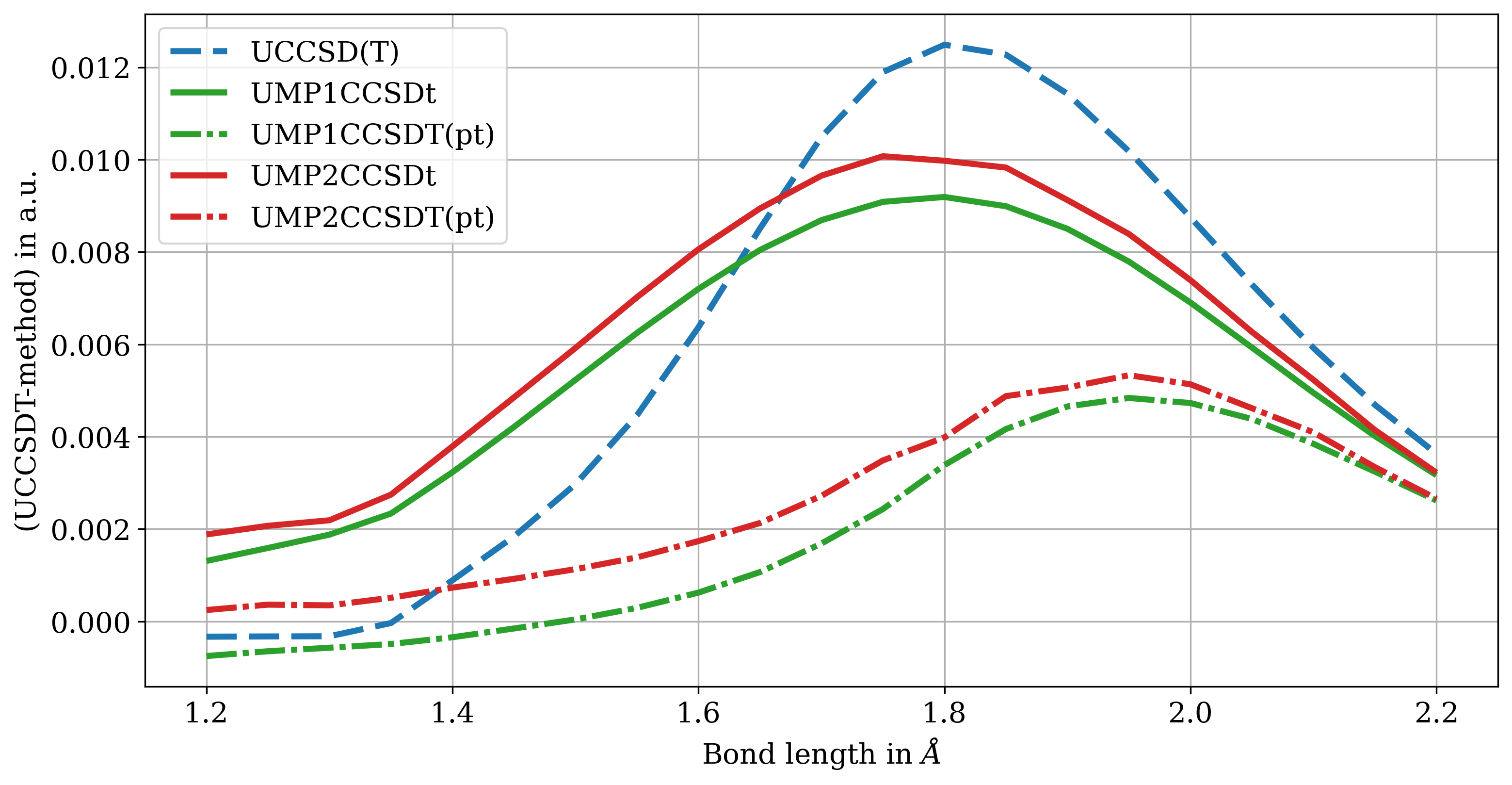}
    \caption{Electronic energy error with respect to UCCSDT along a PEC for F$_2$ in the cc-pCVTZ basis set. AVAS active space is constructed with cc-pVDZ basis set. We set the energy at dissociation as the zero energy for each method.}
    \label{fig:F2_pes}
\end{figure}

Next, we examine the single-bond stretching of the F$_2$ molecule using the potential energy curves shown in Fig. \ref{fig:F2_pes}. Many of the trends observed in the \ce{N_2} study persist in this case. Specifically, (a) the CCSD(T) method performs reasonably well at short bond lengths but becomes increasingly unreliable as the bond is stretched; and (b) employing CCSDT as the fragment solver substantially reduces the NPE to 9 mH. This indicates that the missing dynamical correlation arising from environmental triples is even more critical for F$_2$ than for \ce{N_2}. Indeed, upon including the environmental triples correction, the NPE is further reduced to 4 mH. Nevertheless, we find that incorporating the $T_{3}^{\rm E}$ amplitudes perturbatively is insufficient to fully eliminate the pronounced deviations observed in the bond-length range of 1.6–2.0~\AA. We did not perform UMPCCSDT(it) calculations for F$_2$ in light of the non-smoothness issue encountered above for N$_2$.

\subsection{Total atomization energy comparison:}

We will now discuss the performance of our method by evaluating the total bond dissociation energy (BDE) of a few 3d transition metal hydrides and the total atomization energy (TAE) of three triatomic molecules from the W4-11 dataset. 

\subsubsection{Transition metal hydrides:}


We selected MnH, CrH, CoH and FeH molecules, which are challenging due to the presence of both static and dynamical correlation effects \cite{hait2019levels,TMH_Kohn_JPC2024}, scalar relativistic effects \cite{TMH_Truhlar_JCTC2017, TMH_LanCheng_JCTC2017} and significant basis set effects \cite{TMH_PetersonJCTC2017}. We focus on the electron correlation problem, and thus use the non-relativistic Hamiltonian, with a large enough basis set (cc-pwCVTZ for the transition metals (TMs) and cc-pVDZ for the H atom). We do not freeze any orbitals, as core-valence correlation is significant\cite{TMH_LanCheng_JCTC2017}, and we want to investigate a low-level treatment of the core orbitals via embedding. The active (fragment) space excludes the Ne core of the TM atoms. The def2-SV(P) basis set was chosen within the AVAS scheme to generate fragment orbitals, which includes 5s, 4p, 4d orbitals in addition to the valence atomic orbitals, in order to capture key dynamical correlations with the fragment solver. For the MnH, FeH and CoH molecules, the equilibrium geometry is taken from \citenum{TMH_Kohn_JPC2024}, and the structure of CrH is taken from \citenum{TMH_Truhlar_JCTC2017}. We again employ unrestricted orbitals and cluster amplitudes for all the molecules unless otherwise stated.

The results are summarized in Tables \ref{Table:ccsd_tmh} (singles and doubles only) and \ref{Table:ccsdt_tmh} (with triples). MnH is in the high-spin septet state ($^7\Sigma^{+}$), and Mn is also high-spin ($^6S$). These high-spin configurations should make the electronic structure relatively straightforward: the total triples contribution to the BDE is small ($\sim$ 4 kJ/mol), and CCSD(T) recovers it almost completely. We next consider the embedding schemes, where it turns out that things are not so straightforward. The first-order (MP1) low-level method within the MP1CCSDt scheme reproduces the BDE very well. However, when we improve the embedding by including environment triples, either perturbatively (UMP1CCSDT(pt)) or iteratively (UMP1CCSDT(it)), the agreement deteriorates! It also deteriorates when the (better) second-order (MP2) low-level method is used within the MP2CCSDt scheme. This indicates a fortuitous error cancellation in UMP1CCSDt, which is supported by comparing total energies against CCSDT (Table \ref{tab:mnh_energies_rounded}). Both UMP1CCSDT(pt) and UMP1CCSDT(it) substantially overestimate the correlation energy (by more than 10 mH) for both MnH and Mn.
By contrast, with MP2 environment singles and doubles,  the inclusion of environment triples, via UMP2CCSDT(pt) or (it), reduces the BDE error to less than 0.5 kJ/mol, or about a seven-fold reduction from 3.4 kJ/mol error in UMP2CCSDt. We conclude that second-order treatment of environment singles and doubles is necessary for even the seemingly easy case of MnH.

\begin{table}[h!]
\centering
\caption{Bond Dissociation Energies (BDE) of the transition metal hydrides at the CCSD level in kJ/mol. We have compared two variants of the low-level theory, denoted as MP1 and MP2. BDE is reported at the CCSD level, and for the approximate methods, the differences w.r.t.~CCSD were provided.}
\begin{tabular}{lcccc}
\hline
Method & MnH & CrH & CoH & FeH \\
\hline
CCSD	  & 152.90 & 192.32 & 236.23 &	252.80 \\
UMP1CCSD  &	  3.75 &  -0.10	&  16.66 &	14.04   \\
UMP2CCSD  &  -0.21 &  -0.07	&  -1.47 &	-2.24   \\

\hline
\end{tabular}
\label{Table:ccsd_tmh}
\end{table}

\begin{table}[h!]
\centering
\caption{Bond Dissociation Energies (BDE) of the transition metal hydrides at the CCSDT level in kJ/mol. Two variants of the low-level method (MP1 and MP2) and a few variants of the embedded triples method reported in Sec. \ref{Sec:theory_fragsolver} were compared. BDE is reported at the CCSDT level, and for the approximate methods, the errors w.r.t. CCSDT were provided. The large basis set used is cc-pwCVTZ, and the small basis set used to construct the active space is def2-sv(p). We have applied a Ne core when the fragment space is constructed.}
\begin{tabular}{lcccc}
\hline
Method & MnH & CrH & CoH & FeH \\
\hline
CCSDT &	148.81 & 201.72 & 247.34 & 256.03 \\
CCSD(T) & 0.54 & 7.42 &	35.83 & 14.91 \\
UMP1CCSDt &	0.71 & 6.21 & 22.42	& 13.73 \\
UMP1CCSDT(pt) &	4.77 &	0.72 & 33.06 & 23.92 \\
UMP1CCSDT(it) &	3.49 &	1.36 & 14.98 &	18.71 \\
UMP2CCSDt     &	-3.40 &	4.23 & 4.69	 & -2.31 \\
UMP2CCSDT(pt) &	0.32 &	-0.93 &	6.04 &	1.63 \\
UMP2CCSDT(it) &	-0.47 &	0.58 & 2.24	 &  0.85 \\
\hline
\end{tabular}
\label{Table:ccsdt_tmh}
\end{table}

Next, we consider the $^6\Sigma^+$ state of the CrH molecule, which comprises five unpaired parallel spins and one pair of coupled spins in the valence shell. As a result, CrH exhibits a more complex electronic structure than MnH. This complexity is reflected in the total BDE, which at the CCSD level is approximately 9 kJ/mol lower than that obtained with the CCSDT method. The CCSD(T) method recovers only about 2 kJ/mol of the triples correction, which is not satisfactory. When we apply the embedding method with an MP1-level treatment of the environment SD amplitudes, the total BDE is reproduced more accurately at the UMP1CCSDt level than with CCSD(T). By including the environment triples contribution, either perturbatively or iteratively, we achieve nearly the same accuracy as CCSDT ($\sim$ 1 kJ/mol error). Unsurprisingly, using the improved UMP2CCSDT methods leads to only marginal improvements in the BDE. Overall, CrH represents a relatively straightforward case for the embedding methods, and all variants show satisfactory performance.

FeH has a $^4\Delta$ ground state, with three unpaired parallel spins and three pairs of coupled spins in the valence shell. Although the triples contribution to the BDE is relatively small, $\sim$ 3.2 kJ/mol, FeH exhibits interesting behavior when approximate methods are applied. For example, the deviation of CCSD(T) from CCSDT is significantly larger (15 kJ/mol) than that of CCSD. This is because the perturbative triples correction cannot restore the broken spin symmetry introduced by the UHF reference. In contrast, when the BDE is evaluated using a restricted open-shell Hartree–Fock (ROHF) reference, the difference between CCSDT and CCSD(T) is reduced more than three-fold to $\sim$ 4 kJ/mol. Turning to the embedding approaches, employing MP1 as the low-level treatment for the SD amplitudes leads to a significant deviation at the UMP1CCSD level. This deviation is recovered only when the improved perturbative method, UMP2CCSD, is used as the low-level treatment. A similar trend is observed when comparing the triples variants of the embedding methods. Notably, the second-order variant of the low-level SD treatment performs consistently well across all triples variants, in a way that is similar to what we saw with MnH.

CoH ($^3\Phi$ ground state) is the most challenging of our 4 TM hydrides. The 11 kJ/mol triples contribution is three times smaller than the CCSD(T) error (36 kJ/mol) with UHF orbitals. This is a more drastic version of the failure of CCSD(T) discussed above for FeH, and again can be remedied with an ROHF reference, which reduces the CCSD(T) error to $\sim$ 0.5 kJ/mol. Like FeH, we find that the second-order variants (UMP2CCSDt and UMP2CCSDT) perform much better than their first-order cousins. Unlike FeH, however, the iterative evaluation of $T_{3}^{\rm E}$ is significantly ($\sim$ 4 kJ/mol) more accurate than the corresponding perturbative treatment.

\subsubsection{Strongly correlated systems from the W4-11 dataset:}

We analyzed the total atomization energy (TAE) of three triatomic molecules from the W4-11 dataset, FO$_2$, O$_3$, and NO$_2$. For these molecules, the improvement obtained with CCSD(T) is so exceptionally large ($\sim$15 kcal/mol) that they are often regarded as multireference systems even in their ground states. We employed the cc-pCVTZ basis set as the large basis for all atoms and the cc-pVDZ basis set as the minimal basis to construct the active space using the AVAS scheme. In our earlier work \cite{mpcc_jcp24}, we showed that an MP1-level treatment of the environment SD amplitudes is sufficient to recover TAE within chemical accuracy for the UMPCCSD method.

In this work, we compared a few variants of the embedding method combined with the triples solver (Table \ref{Table:w411_tae}), where the low-level method corresponds to the two variants discussed in Sec.~\ref{Sec:theory_lowlevel}. For all three molecules, both low-level methods perform equally well with the SDT solver, consistent with the behavior observed for the SD solver. Apparently, the complexities of the TM hydrides are not an issue here. Even when we neglect the environment triples contribution within the embedding approach, we obtain an accuracy comparable to that of the CCSD(T) method, except for the NO$_2$ molecule. In that case, CCSD(T) performs so well as to be nearly unbeatable, while the UMP2CCSDT embedding methods reduces the CCSD(T) error by more than a factor of ten for \ce{FO2} and \ce{O3}. We note that our (UCCSDT-UCCSD(T)) difference is much larger than that previously reported \cite{W411_Karton_MartinJPC2024}, probably because we retain the full core and not resort to composite approaches to obtain the UCCSDT energies.  

\begin{table}[h!]
\centering
\caption{Total Atomization Energies (TAE) in kcal/mol for FO$_2$, O$_3$ and NO$_2$ molecules. Two variants of the low-level method (MP1 and MP2) and a few variants of the embedded triples method reported in Sec. \ref{Sec:theory_fragsolver} were compared. TAE is reported at the CCSDT level, and for the approximate methods and CCSD, the errors w.r.t. CCSDT were provided. The large basis set used is cc-pCVTZ, and the small basis set used to construct the active space is cc-pVDZ.}
\begin{tabular}{lccc}
\hline
Method & FO$_2$ & O$_3$ & NO$_2$ \\
\hline
UCCSDT                  & 121.83 & 132.36   & 212.83 \\
UCCSD                   & 18.93	&	21.35	&	17.56 \\
UCCSD(T)                &  4.40	&	5.41	&	-0.51  \\
UMP1CCSDt               &  5.35	&	5.73	&	3.33  \\
UMP1CCSDT(pt)           & -0.51	&	-0.62	&	-2.60 \\
UMP2CCSDt               &  5.92	&	7.17	&	3.47 \\
UMP2CCSDT(pt)           &  0.23	&	0.43	&	-1.58 \\
\hline
\end{tabular}
\label{Table:w411_tae}
\end{table}

\section{Conclusion}

We have developed, implemented, and numerically explored several variants of a triples extension of our static embedding scheme, MPCC \cite{mpcc_jcp24}. These MPCC embedding methods use a coupled cluster singles, doubles, and triples (CCSDT) solver in an active (fragment) region, coupled to a lower order M{\o}ller-Plesset (MP) perturbative treatment of the environment. This higher-accuracy CCSDT solver enabled us to achieve both qualitative and quantitative accuracy beyond the CCSD and CCSD(T) level on various problems. Examples of such applications include radical chemistry and transition metal chemistry.

However, the inclusion of a CCSDT solver introduces the challenge that the feedback from the environment must be adequately captured when triples are introduced in the fragment. Within the MPCCSD scheme, we captured the feedback of the (first order) environment SD amplitudes on the fragment SD amplitudes to all orders of perturbation, while the feedback of the fragment amplitudes on the environment amplitudes was included at first order. Unfortunately, treating environment triples on the same footing as environment doubles is not feasible due to the steep scaling of triples in both memory and computational cost. 

Instead, as summarized in Table \ref{tab:compute-memory}, we explored three approximate schemes to handle the environment triples, and their influence on the active fragment. These schemes involve macroiterations to update the environment amplitudes, microiterations to update the fragment amplitudes, and (sometimes) a final non-iterative step.
\begin{enumerate}
    \item The simplest possible approach is to completely neglect the environment triples, leading to  MPCCSDt, where the lowercase ``t'' indicates that only fragment triples are included.
    \item The next simplest approach is MPCCSDT(pt), wherein we neglect the influence of environment triples on the fragment amplitudes. The environment triples are described by a second-order amplitude equation, in which the fragment singles, doubles and triples amplitudes all contribute to the environment triples (neglecting the $[V, T_3]$ terms to avoid their $\mathcal{O}(V^5 O^3)$ computational cost). The resulting equations are solved by semicanonicalizing the orbital basis.   
    \item In the most advanced approach, MPCCSDT(it), we also account for the influence of environment triples on the fragment amplitudes, which modifies the fragment Hamiltonian through an additional similarity transformation by the environment triples. This procedure incurs a macro iterative $\mathcal{O}(N^7)$ cost.
\end{enumerate}
Each of these 3 methods can evaluate the SD amplitudes using first-order low-level equations (MP1) as used in our earlier work, or, at additional compute cost, by also including second-order terms in the resulting MP2 equations. The MP2 equations are consistent in terms of the perturbative order with the low-level treatment used for the environment triples amplitudes. Thus, we have a total of 6 candidate embedded triples methods (Table \ref{tab:compute-memory}).

The 6 different candidate MPCCSDT methods were then numerically tested on: a) bond-stretching of N$_2$ and F$_2$; b) TAEs of three multi-reference systems: FO$_2$, O$_3$, and NO$_2$; c) BDE analysis of 4 transition metal hydrides: MnH, CrH, FeH and CoH. The fragment was defined by a medium-sized basis set (cc-pVDZ) with a neon core excluded for the transition metals, while the environment contained any frozen orbitals and additional orbitals from a larger basis set (cc-pwCVTZ for TMs, and otherwise cc-pCVTZ). Unrestricted orbitals were used throughout. The major findings were:
\begin{enumerate}
    \item The neglect of environment triples via either MP1CCSDt or MP2CCSDt typically leads to unacceptably large errors, and therefore cannot be recommended.
    \item Generally, we recommend the MP2CCSDT(pt) approach over MP1CCSDT(pt). MP2CCSDT(pt) almost always significantly outperforms CCSD(T), often reducing CCSD(T) errors in strongly correlated cases by over a factor of 10, while having the same rate-determining compute cost in the limit of a large environment. 
    \item In the most strongly correlated molecules, such as CoH, FeH, the use of second order environment amplitudes is important, and MP2CCSDT(pt) significantly outperforms MP1CCSDT(pt). In other cases the difference is smaller. CoH and FeH are also the only cases we examined where MP2CCSDT(it) significantly improves upon MP2CCSDT(pt), with up to 2.5 times error reduction.
\end{enumerate}

    


Although we have identified a systematic way to improve the accuracy of the UMPCCSDT method, several aspects require further development. Improved scaling could be achieved by adopting linear scaling methods for the perturbative environment amplitudes\cite{wang2025more}, or conceivably also the fragment solver. 
Many strong correlation problems, such as the ones involving multiple metal centres, for example, cubane, Fe-Mo cofactor of the nitrogenase enzyme, may need a solver that incorporates higher-order correlations than CCSDT. To this end, we aim to incorporate a selected configuration interaction (sCI) \cite{ACI_Evangelista_JCP2014, SHCI_UmrigarJCTC2016, SCI_Scemama_Caffarel_JCP2017, ASCI_Tubman_JCTC2020, cotton2022truncated} method and quantum hardware-inspired quantum subspace diagonalization (QSD) \cite{QSCI_NakagawaJCTC2024, SQD_IBMScience2025} method as a solver in the future.

\section{Acknowledgements}

This material is based upon work supported by the U.S. Department of Energy, Office of Science, Office of Advanced Scientific Computing Research and Office of Basic Energy Sciences, Scientific Discovery through Advanced Computing (SciDAC) program under Award Number DE‐SC0022198 (A.S., K.B.W. and L.L.), and Award Number DE-SC0022364 (M.H-G.). This research used resources of the National Energy Research Scientific Computing Center, a DOE Office of Science User Facility supported by the Office of Science of the U.S. Department of Energy under Contract No. DE-AC02- 05CH11231 using NERSC award BES-ERCAP0029462.

\bibliographystyle{apsrev4-1}
\bibliography{coupled_cluster, embedding, misc}

@article{Cizek_CC,
author = {\ifmmode \check{C}\else \v{C}\fi{}\'{\i}\ifmmode \check{z}\else \v{z}\fi{}ek, J.},
title = {On the Correlation Problem in Atomic and Molecular Systems. Calculation of Wavefunction Components in Ursell-Type Expansion Using Quantum-Field Theoretical Methods},
journal = {J. Chem. Phys.},
volume = {45},
number = {11},
pages = {4256-4266},
year = {1966},
doi = {10.1063/1.1727484},
URL = { http://dx.doi.org/10.1063/1.1727484},
eprint = {http://dx.doi.org/10.1063/1.1727484}
}

@article{Paldus_Cizek,
  title = {Correlation Problems in Atomic and Molecular Systems. IV. Extended Coupled-Pair Many-Electron Theory and Its Application to the B${\mathrm{H}}_{3}$ Molecule},
  author = {Paldus, J. and \ifmmode \check{C}\else \v{C}\fi{}\'{\i}\ifmmode \check{z}\else \v{z}\fi{}ek, J. and Shavitt, I.},
  journal = {Phys. Rev. A},
  volume = {5},
  issue = {1},
  pages = {50--67},
  numpages = {0},
  year = {1972},
  month = {Jan},
  publisher = {American Physical Society},
  doi = {10.1103/PhysRevA.5.50},
  url = {http://link.aps.org/doi/10.1103/PhysRevA.5.50}
}

@article{cc_downfold_kowalski_JCP18,
    author = {Kowalski, Karol},
    title = {Properties of coupled-cluster equations originating in excitation sub-algebras},
    journal = {The Journal of Chemical Physics},
    volume = {148},
    number = {9},
    pages = {094104},
    year = {2018},
    month = {03},
    issn = {0021-9606},
    doi = {10.1063/1.5010693},
    url = {https://doi.org/10.1063/1.5010693}
}

@article{cc_downfold_kowalski23,
    author = {Kowalski, Karol},
    title = "{Sub-system self-consistency in coupled cluster theory}",
    journal = {J. Chem. Phys.},
    volume = {158},
    number = {5},
    pages = {054101},
    year = {2023},
    month = {02},
    doi = {10.1063/5.0125696},
    url = {https://doi.org/10.1063/5.0125696}
}

@article{cc_downfold_karol19,
    author = {Bauman, Nicholas P. and Bylaska, Eric J. and Krishnamoorthy, Sriram and Low, Guang Hao and Wiebe, Nathan and Granade, Christopher E. and Roetteler, Martin and Troyer, Matthias and Kowalski, Karol},
    title = "{Downfolding of many-body Hamiltonians using active-space models: Extension of the sub-system embedding sub-algebras approach to unitary coupled cluster formalisms}",
    journal = {J. Chem. Phys.},
    volume = {151},
    number = {1},
    pages = {014107},
    year = {2019},
    month = {07},
    doi = {10.1063/1.5094643},
    url = {https://doi.org/10.1063/1.5094643}
}

@article{MLCCJCP2014,
    author = {Myhre, Rolf H. and Sánchez de Merás, Alfredo M. J. and Koch, Henrik},
    title = "{Multi-level coupled cluster theory}",
    journal = {J. Chem. Phys.},
    volume = {141},
    number = {22},
    pages = {224105},
    year = {2014},
    month = {12},
    issn = {0021-9606},
    doi = {10.1063/1.4903195},
    url = {https://doi.org/10.1063/1.4903195}
}

@article{MLCCjctc2021,
author = {Folkestad, Sarai Dery and Kj{\o}nstad, Eirik F. and Goletto, Linda and Koch, Henrik},
title = {Multilevel CC2 and CCSD in Reduced Orbital Spaces: Electronic Excitations in Large Molecular Systems},
journal = {J. Chem. Theory Comput.},
volume = {17},
number = {2},
pages = {714-726},
year = {2021},
doi = {10.1021/acs.jctc.0c00590},
URL = {https://doi.org/10.1021/acs.jctc.0c00590    
}
}

@article{CCt3_JCTCShen2012,
author = {Shen, Jun and Piecuch, Piotr},
title = {Merging Active-Space and Renormalized Coupled-Cluster Methods via the CC(P;Q) Formalism, with Benchmark Calculations for Singlet–Triplet Gaps in Biradical Systems},
journal = {J. Chem. Theory Comput.},
volume = {8},
number = {12},
pages = {4968-4988},
year = {2012},
doi = {10.1021/ct300762m},
URL = {https://doi.org/10.1021/ct300762m},
eprint = {https://doi.org/10.1021/ct300762m}
}

@article{CCSD_TMHGCPL89,
title = {A fifth-order perturbation comparison of electron correlation theories},
journal = {Chem. Phys. Lett.},
volume = {157},
number = {6},
pages = {479-483},
year = {1989},
issn = {0009-2614},
doi = {https://doi.org/10.1016/S0009-2614(89)87395-6},
url = {https://www.sciencedirect.com/science/article/pii/S0009261489873956},
author = {Krishnan Raghavachari and Gary W. Trucks and John A. Pople and Martin Head-Gordon},
}

@article{ccsdt3nogaJCP87,
    author = {Noga, Jozef and Bartlett, Rodney J.},
    title = {The full CCSDT model for molecular electronic structure},
    journal = {The Journal of Chemical Physics},
    volume = {86},
    number = {12},
    pages = {7041-7050},
    year = {1987},
    month = {06},
    issn = {0021-9606},
    doi = {10.1063/1.452353},
    url = {https://doi.org/10.1063/1.452353}
}

@article{ShenLi_CCSDt_JCP2010,
    author = {Shen, Jun and Xu, Enhua and Kou, Zhuangfei and Li, Shuhua},
    title = {A coupled cluster approach with a hybrid treatment of connected triple excitations for bond-breaking potential energy surfaces},
    journal = {The Journal of Chemical Physics},
    volume = {132},
    number = {11},
    pages = {114115},
    year = {2010},
    month = {03},
    issn = {0021-9606},
    doi = {10.1063/1.3359851}
}

@article{MMCC_Piecuch01102002,
author = {Piotr Piecuch and Karol Kowalski and Ian S. O. Pimienta and Michael J. Mcguire},
title = {Recent advances in electronic structure theory: Method of moments of coupled-cluster equations and renormalized coupled-cluster approaches},
journal = {International Reviews in Physical Chemistry},
volume = {21},
number = {4},
pages = {527--655},
year = {2002},
publisher = {Taylor \& Francis},
doi = {10.1080/0144235021000053811},
URL ={https://doi.org/10.1080/0144235021000053811
}
}

@article{schafer_gruneis_NatComm2025,
  title={Understanding discrepancies in noncovalent interaction energies from wavefunction theories for large molecules},
  author={Sch{\"a}fer, Tobias and Irmler, Andreas and Gallo, Alejandro and Gr{\"u}neis, Andreas},
  journal={Nature Communications},
  volume={16},
  number={1},
  pages={9108},
  year={2025},
  doi = {10.1038/s41467-025-64104-8},  
  publisher={Nature Publishing Group UK London}
}

@article{Schafer_Gruneis_PhysRevLett2023,
  title = {Averting the Infrared Catastrophe in the Gold Standard of Quantum Chemistry},
  author = {Masios, Nikolaos and Irmler, Andreas and Sch\"afer, Tobias and Gr\"uneis, Andreas},
  journal = {Phys. Rev. Lett.},
  volume = {131},
  issue = {18},
  pages = {186401},
  numpages = {6},
  year = {2023},
  month = {Oct},
  publisher = {American Physical Society},
  doi = {10.1103/PhysRevLett.131.186401},
  url = {https://link.aps.org/doi/10.1103/PhysRevLett.131.186401}
}

@article{Tkatchenko_Nagy_NatComm2021,
  title={Interactions between large molecules pose a puzzle for reference quantum mechanical methods},
  author={Al-Hamdani, Yasmine S and Nagy, P{\'e}ter R and Zen, Andrea and Barton, Dennis and K{\'a}llay, Mih{\'a}ly and Brandenburg, Jan Gerit and Tkatchenko, Alexandre},
  journal={Nature Communications},
  volume={12},
  number={1},
  pages={3927},
  year={2021},
  doi = {10.1038/s41467-021-24119-3},
  publisher={Nature Publishing Group UK London}
}

@article{Bartlett_JCP98,
    author = {Kucharski, Stanisl/aw A. and Bartlett, Rodney J.},
    title = {An efficient way to include connected quadruple contributions into the coupled cluster method},
    journal = {The Journal of Chemical Physics},
    volume = {108},
    number = {22},
    pages = {9221-9226},
    year = {1998},
    month = {06},    
    issn = {0021-9606},
    doi = {10.1063/1.476376},
    url = {https://doi.org/10.1063/1.476376}
}

@article{Georges96,
  title = {Dynamical mean-field theory of strongly correlated fermion systems and the limit of infinite dimensions},
  author = {Georges, Antoine and Kotliar, Gabriel and Krauth, Werner and Rozenberg, Marcelo J.},
  journal = {Rev. Mod. Phys.},
  volume = {68},
  issue = {1},
  pages = {13--125},
  numpages = {0},
  year = {1996},
  month = {Jan},
  publisher = {American Physical Society},
  doi = {10.1103/RevModPhys.68.13},
  url = {https://link.aps.org/doi/10.1103/RevModPhys.68.13}
}

@article{Zhu_DMFT_PRX21,
  title = {Ab Initio Full Cell $GW+\mathrm{DMFT}$ for Correlated Materials},
  author = {Zhu, Tianyu and Chan, Garnet Kin-Lic},
  journal = {Phys. Rev. X},
  volume = {11},
  issue = {2},
  pages = {021006},
  numpages = {13},
  year = {2021},
  month = {Apr},
  publisher = {American Physical Society},
  doi = {10.1103/PhysRevX.11.021006},
  url = {https://link.aps.org/doi/10.1103/PhysRevX.11.021006}
}

@article{DMFT_infinite_dim_Georges_Kotliar_1992,
  title = {Hubbard model in infinite dimensions},
  author = {Georges, Antoine and Kotliar, Gabriel},
  journal = {Phys. Rev. B},
  volume = {45},
  issue = {12},
  pages = {6479--6483},
  numpages = {0},
  year = {1992},
  month = {Mar},
  publisher = {American Physical Society},
  doi = {10.1103/PhysRevB.45.6479},
  url = {https://link.aps.org/doi/10.1103/PhysRevB.45.6479}
}

@article{DMET_bootstrap_jcp_2016,
author = {Matthew Welborn and Takashi Tsuchimochi and Troy Van Voorhis},
title = {Bootstrap embedding: An internally consistent fragment-based method},
journal = {J. Chem. Phys.},
volume = {145},
number = {7},
pages = {074102},
year = {2016},
doi = {10.1063/1.4960986},
URL = {http://dx.doi.org/10.1063/1.4960986},
eprint = {http://dx.doi.org/10.1063/1.4960986}
}

@article{Knizia12_DMET,
  title = {Density Matrix Embedding: A Simple Alternative to Dynamical Mean-Field Theory},
  author = {Knizia, Gerald and Chan, Garnet Kin-Lic},
  journal = {Phys. Rev. Lett.},
  volume = {109},
  issue = {18},
  pages = {186404},
  numpages = {5},
  year = {2012},
  month = {Nov},
  publisher = {American Physical Society},
  doi = {10.1103/PhysRevLett.109.186404},
  url = {https://link.aps.org/doi/10.1103/PhysRevLett.109.186404}
}

@article{Knizia13_DMET,
author = {Knizia, Gerald and Chan, Garnet Kin-Lic},
title = {Density Matrix Embedding: A Strong-Coupling Quantum Embedding Theory},
journal = {J. Chem. Theory Comput.},
volume = {9},
number = {3},
pages = {1428-1432},
year = {2013},
doi = {10.1021/ct301044e},
note ={PMID: 26587604},
URL = {https://doi.org/10.1021/ct301044e}
}

@article{Zhu_PRB19,
  title = {Coupled-cluster impurity solvers for dynamical mean-field theory},
  author = {Zhu, Tianyu and Jim\'enez-Hoyos, Carlos A. and McClain, James and Berkelbach, Timothy C. and Chan, Garnet Kin-Lic},
  journal = {Phys. Rev. B},
  volume = {100},
  issue = {11},
  pages = {115154},
  numpages = {9},
  year = {2019},
  month = {Sep},
  publisher = {American Physical Society},
  doi = {10.1103/PhysRevB.100.115154},
  url = {https://link.aps.org/doi/10.1103/PhysRevB.100.115154}
}

@article{mpcc_jcp24,
    author = {Shee, Avijit and Faulstich, Fabian M. and Whaley, K. Birgitta and Lin, Lin and Head-Gordon, Martin},
    title = {A static quantum embedding scheme based on coupled cluster theory},
    journal = {The Journal of Chemical Physics},
    volume = {161},
    number = {16},
    pages = {164107},
    year = {2024},
    month = {10}, 
    issn = {0021-9606},
    doi = {10.1063/5.0214065},
    url = {https://doi.org/10.1063/5.0214065}
}

@article{sun2020recent,
  title={Recent developments in the PySCF program package},
  author={Sun, Qiming and Zhang, Xing and Banerjee, Samragni and Bao, Peng and Barbry, Marc and Blunt, Nick S and Bogdanov, Nikolay A and Booth, George H and Chen, Jia and Cui, Zhi-Hao and others},
  journal={The Journal of chemical physics},
  volume={153},
  number={2},
  year={2020},
  publisher={AIP Publishing}
}

@article{bochevarov2006hybrid,
  title={Hybrid correlation models based on active-space partitioning: Seeking accurate O (N5) ab initio methods for bond breaking},
  author={Bochevarov, Arteum D and Temelso, Berhane and Sherrill, C David},
  journal={The Journal of chemical physics},
  volume={125},
  number={5},
  year={2006},
  publisher={AIP Publishing}
}

@article{bochevarov2005hybrid,
  title={Hybrid correlation models based on active-space partitioning: Correcting second-order M{\o}ller--Plesset perturbation theory for bond-breaking reactions},
  author={Bochevarov, Arteum D and Sherrill, C David},
  journal={The Journal of chemical physics},
  volume={122},
  number={23},
  year={2005},
  publisher={AIP Publishing}
}

@article{nooijen1999combining,
  title={Combining coupled cluster and perturbation theory},
  author={Nooijen, Marcel},
  journal={The Journal of Chemical Physics},
  volume={111},
  number={24},
  pages={10815--10826},
  year={1999},
  publisher={American Institute of Physics}
}

@article{adamowicz1998state,
  title={The state-selective coupled cluster method for quasi-degenerate electronic states},
  author={Adamowicz, Ludwik and Piecuch, Piotr and Ghose, Keya B},
  journal={Molecular Physics},
  volume={94},
  number={1},
  pages={225--234},
  year={1998},
  publisher={Taylor \& Francis}
}

@article{piecuch1994state,
  title={State-selective multireference coupled-cluster theory employing the single-reference formalism: Implementation and application to the H8 model system},
  author={Piecuch, Piotr and Adamowicz, Ludwik},
  journal={The Journal of chemical physics},
  volume={100},
  number={8},
  pages={5792--5809},
  year={1994},
  publisher={American Institute of Physics}
}

@article{piecuch1993state,
  title={A state-selective multireference coupled-cluster theory employing the single-reference formalism},
  author={Piecuch, Piotr and Oliphant, Nevin and Adamowicz, Ludwik},
  journal={The Journal of chemical physics},
  volume={99},
  number={3},
  pages={1875--1900},
  year={1993},
  publisher={AIP Publishing}
}

@article{Shee2019,
  doi = {10.1021/acs.jctc.9b00603},
  url = {https://doi.org/10.1021/acs.jctc.9b00603},
  year = {2019},
  month = sep,
  publisher = {American Chemical Society ({ACS})},
  volume = {15},
  number = {11},
  pages = {6010--6024},
  author = {Avijit Shee and Dominika Zgid},
  title = {Coupled Cluster as an Impurity Solver for Green's Function Embedding Methods},
  journal = {J. Chem. Theory Comput.}
}

@article{AVASknizia2017,
author = {Sayfutyarova, Elvira R. and Sun, Qiming and Chan, Garnet Kin-Lic and Knizia, Gerald},
title = {Automated Construction of Molecular Active Spaces from Atomic Valence Orbitals},
journal = {J. Chem. Theory Comput.},
volume = {13},
number = {9},
pages = {4063-4078},
year = {2017},
doi = {10.1021/acs.jctc.7b00128},    
URL = {https://doi.org/10.1021/acs.jctc.7b00128}
}

@article{W4_KARTON_CPL2011,
title = {W4-11: A high-confidence benchmark dataset for computational thermochemistry derived from first-principles W4 data},
journal = {Chem. Phys. Lett.},
volume = {510},
number = {4},
pages = {165-178},
year = {2011},
issn = {0009-2614},
doi = {https://doi.org/10.1016/j.cplett.2011.05.007},
url = {https://www.sciencedirect.com/science/article/pii/S0009261411005616},
author = {Amir Karton and Shauli Daon and Jan M.L. Martin}
}

@article{sun2018pyscf,
  title={PySCF: the Python-based simulations of chemistry framework},
  author={Sun, Qiming and Berkelbach, Timothy C and Blunt, Nick S and Booth, George H and Guo, Sheng and Li, Zhendong and Liu, Junzi and McClain, James D and Sayfutyarova, Elvira R and Sharma, Sandeep and others},
  journal={Wiley Interdisciplinary Reviews: Computational Molecular Science},
  volume={8},
  number={1},
  pages={e1340},
  year={2018},
  publisher={Wiley Online Library}
}

@article{sun2015libcint,
  title={Libcint: An efficient general integral library for gaussian basis functions},
  author={Sun, Qiming},
  journal={Journal of computational chemistry},
  volume={36},
  number={22},
  pages={1664--1671},
  year={2015},
  publisher={Wiley Online Library}
}

@article{thouless1960stability,
  title={Stability conditions and nuclear rotations in the Hartree-Fock theory},
  author={Thouless, David J},
  journal={Nuclear Physics},
  volume={21},
  pages={225--232},
  year={1960},
  publisher={Elsevier}
}

@article{kurlancheek2009violations,
  title={Violations of N-representability from spin-unrestricted orbitals in M{\o}ller--Plesset perturbation theory and related double-hybrid density functional theory},
  author={Kurlancheek, Westin and Head-Gordon, Martin},
  journal={Mol. Phys.},
  volume={107},
  number={8-12},
  pages={1223--1232},
  year={2009},
  publisher={Taylor \& Francis}
}

@article{TMH_PetersonJCTC2017,
author = {Fang, Zongtang and Vasiliu, Monica and Peterson, Kirk A. and Dixon, David A.},
title = {Prediction of Bond Dissociation Energies/Heats of Formation for Diatomic Transition Metal Compounds: CCSD(T) Works},
journal = {Journal of Chemical Theory and Computation},
volume = {13},
number = {3},
pages = {1057-1066},
year = {2017},
doi = {10.1021/acs.jctc.6b00971},
    note ={PMID: 28080051},
URL = {https://doi.org/10.1021/acs.jctc.6b00971}
}

@article{TMH_Truhlar_JCTC2017,
author = {Xu, Xuefei and Zhang, Wenjing and Tang, Mingsheng and Truhlar, Donald G.},
title = {Do Practical Standard Coupled Cluster Calculations Agree Better than Kohn–Sham Calculations with Currently Available Functionals When Compared to the Best Available Experimental Data for Dissociation Energies of Bonds to 3d Transition Metals?},
journal = {Journal of Chemical Theory and Computation},
volume = {11},
number = {5},
pages = {2036-2052},
year = {2015},
doi = {10.1021/acs.jctc.5b00081},
    note ={PMID: 26574408},
URL = {https://doi.org/10.1021/acs.jctc.5b00081}
}

@article{TMH_LanCheng_JCTC2017,
author = {Cheng, Lan and Gauss, J{\"u}rgen and Ruscic, Branko and Armentrout, Peter B. and Stanton, John F.},
title = {Bond Dissociation Energies for Diatomic Molecules Containing 3d Transition Metals: Benchmark Scalar-Relativistic Coupled-Cluster Calculations for 20 Molecules},
journal = {Journal of Chemical Theory and Computation},
volume = {13},
number = {3},
pages = {1044-1056},
year = {2017},
doi = {10.1021/acs.jctc.6b00970},
    note ={PMID: 28080054},
URL = {https://doi.org/10.1021/acs.jctc.6b00970}
}

@article{TMH_Kohn_JPC2024,
author = {Waigum, Alexander and Ert{\"u}rk, Murat and K{\"o}hn, Andreas},
title = {Accurate Thermochemistry with Multireference Methods: A Stress Test for Internally Contracted Multireference Coupled-Cluster Theory},
journal = {The Journal of Physical Chemistry A},
volume = {128},
number = {46},
pages = {10053-10070},
year = {2024},
doi = {10.1021/acs.jpca.4c05819},
    note ={PMID: 39535968},
URL = {https://doi.org/10.1021/acs.jpca.4c05819}
}

@article{W411_Karton_MartinJPC2024,
author = {Semidalas, Emmanouil and Karton, Amir and Martin, Jan M. L.},
title = {W4Λ: Leveraging Λ Coupled-Cluster for Accurate Computational Thermochemistry Approaches},
journal = {The Journal of Physical Chemistry A},
volume = {128},
number = {9},
pages = {1715-1724},
year = {2024},
doi = {10.1021/acs.jpca.3c08158},
URL = {https://doi.org/10.1021/acs.jpca.3c08158
}
}

@article{SHCI_UmrigarJCTC2016,
author = {Holmes, Adam A. and Tubman, Norm M. and Umrigar, C. J.},
title = {Heat-Bath Configuration Interaction: An Efficient Selected Configuration Interaction Algorithm Inspired by Heat-Bath Sampling},
journal = {Journal of Chemical Theory and Computation},
volume = {12},
number = {8},
pages = {3674-3680},
year = {2016},
doi = {10.1021/acs.jctc.6b00407},
URL = {https://doi.org/10.1021/acs.jctc.6b00407
}
}

@article{SCI_Scemama_Caffarel_JCP2017,
    author = {Garniron, Yann and Scemama, Anthony and Loos, Pierre-François and Caffarel, Michel},
    title = {Hybrid stochastic-deterministic calculation of the second-order perturbative contribution of multireference perturbation theory},
    journal = {The Journal of Chemical Physics},
    volume = {147},
    number = {3},
    pages = {034101},
    year = {2017},
    month = {07},
    issn = {0021-9606},
    doi = {10.1063/1.4992127},
    url = {https://doi.org/10.1063/1.4992127}
}

@article{ASCI_Tubman_JCTC2020,
author = {Tubman, Norm M. and Freeman, C. Daniel and Levine, Daniel S. and Hait, Diptarka and Head-Gordon, Martin and Whaley, K. Birgitta},
title = {Modern Approaches to Exact Diagonalization and Selected Configuration Interaction with the Adaptive Sampling CI Method},
journal = {Journal of Chemical Theory and Computation},
volume = {16},
number = {4},
pages = {2139-2159},
year = {2020},
doi = {10.1021/acs.jctc.8b00536},
URL = {https://doi.org/10.1021/acs.jctc.8b00536
}
}

@article{ACI_Evangelista_JCP2014,
    author = {Evangelista, Francesco A.},
    title = {Adaptive multiconfigurational wave functions},
    journal = {The Journal of Chemical Physics},
    volume = {140},
    number = {12},
    pages = {124114},
    year = {2014},
    month = {03},
    issn = {0021-9606},
    doi = {10.1063/1.4869192},
    url = {https://doi.org/10.1063/1.4869192}
}

@article{QSCI_NakagawaJCTC2024,
author = {Nakagawa, Yuya O. and Kamoshita, Masahiko and Mizukami, Wataru and Sudo, Shotaro and Ohnishi, Yu-ya},
title = {ADAPT-QSCI: Adaptive Construction of an Input State for Quantum-Selected Configuration Interaction},
journal = {Journal of Chemical Theory and Computation},
volume = {20},
number = {24},
pages = {10817-10825},
year = {2024},
doi = {10.1021/acs.jctc.4c00846},
URL = {https://doi.org/10.1021/acs.jctc.4c00846}
}

@article{SQD_IBMScience2025,
author = {Javier Robledo-Moreno  and Mario Motta  and Holger Haas  and Ali Javadi-Abhari  and Petar Jurcevic  and William Kirby  and Simon Martiel  and Kunal Sharma  and Sandeep Sharma  and Tomonori Shirakawa  and Iskandar Sitdikov  and Rong-Yang Sun  and Kevin J. Sung  and Maika Takita  and Minh C. Tran  and Seiji Yunoki  and Antonio Mezzacapo },
title = {Chemistry beyond the scale of exact diagonalization on a quantum-centric supercomputer},
journal = {Science Advances},
volume = {11},
number = {25},
pages = {eadu9991},
year = {2025},
doi = {10.1126/sciadv.adu9991},
URL = {https://www.science.org/doi/abs/10.1126/sciadv.adu9991}
}

@article{DUCC_Kowalski_JCTC2020,
author = {Metcalf, Mekena and Bauman, Nicholas P. and Kowalski, Karol and de Jong, Wibe A.},
title = {Resource-Efficient Chemistry on Quantum Computers with the Variational Quantum Eigensolver and the Double Unitary Coupled-Cluster Approach},
journal = {Journal of Chemical Theory and Computation},
volume = {16},
number = {10},
pages = {6165-6175},
year = {2020},
doi = {10.1021/acs.jctc.0c00421},
URL = {https://doi.org/10.1021/acs.jctc.0c00421
}
}

@article{DSRG_downfolding_PRX2023,
  title = {Leveraging Small-Scale Quantum Computers with Unitarily Downfolded Hamiltonians},
  author = {Huang, Renke and Li, Chenyang and Evangelista, Francesco A.},
  journal = {PRX Quantum},
  volume = {4},
  issue = {2},
  pages = {020313},
  numpages = {20},
  year = {2023},
  month = {Apr},
  publisher = {American Physical Society},
  doi = {10.1103/PRXQuantum.4.020313},
  url = {https://link.aps.org/doi/10.1103/PRXQuantum.4.020313}
}

@article{isodesmic_pople_jacs1970,
author = {Hehre, Warren J. and Ditchfield, R. and Radom, L. and Pople, John A.},
title = {Molecular orbital theory of the electronic structure of organic compounds. V. Molecular theory of bond separation},
journal = {Journal of the American Chemical Society},
volume = {92},
number = {16},
pages = {4796-4801},
year = {1970},
doi = {10.1021/ja00719a006},
URL = {https://doi.org/10.1021/ja00719a006
}
}

@article{MP4_FRISCH_CPL1980,
title = {A systematic study of the effect of triple substitutions on the electron correlation energy of small molecules},
journal = {Chemical Physics Letters},
volume = {75},
number = {1},
pages = {66-68},
year = {1980},
issn = {0009-2614},
doi = {https://doi.org/10.1016/0009-2614(80)80465-9},
url = {https://www.sciencedirect.com/science/article/pii/0009261480804659},
author = {Michael J. Frisch and Raghavachari Krishnan and John A. Pople}
}

@article{Lowdin01011956,
author = {Per-Olov Löwdin},
title = {Quantum theory of cohesive properties of solids},
journal = {Advances in Physics},
volume = {5},
number = {17},
pages = {1--171},
year = {1956},
publisher = {Taylor \& Francis},
doi = {10.1080/00018735600101155},
URL = {https://doi.org/10.1080/00018735600101155
}
}

@article{Cr_dimer_expt_JPC93,
author = {Casey, Sean M. and Leopold, Doreen G.},
title = {Negative ion photoelectron spectroscopy of chromium dimer},
journal = {The Journal of Physical Chemistry},
volume = {97},
number = {4},
pages = {816-830},
year = {1993},
doi = {10.1021/j100106a005},
URL = {https://doi.org/10.1021/j100106a005
}}

@article{Peterson_JCP2017,
author = {Fang, Zongtang and Vasiliu, Monica and Peterson, Kirk A. and Dixon, David A.},
title = {Prediction of Bond Dissociation Energies/Heats of Formation for Diatomic Transition Metal Compounds: CCSD(T) Works},
journal = {Journal of Chemical Theory and Computation},
volume = {13},
number = {3},
pages = {1057-1066},
year = {2017},
doi = {10.1021/acs.jctc.6b00971},
URL = {https://doi.org/10.1021/acs.jctc.6b00971
}
}

@article{CCSDT_DFT_orbitals_MHG_JCTC2021,
author = {Bertels, Luke W. and Lee, Joonho and Head-Gordon, Martin},
title = {Polishing the Gold Standard: The Role of Orbital Choice in CCSD(T) Vibrational Frequency Prediction},
journal = {Journal of Chemical Theory and Computation},
volume = {17},
number = {2},
pages = {742-755},
year = {2021},
doi = {10.1021/acs.jctc.0c00746},
URL = { https://doi.org/10.1021/acs.jctc.0c00746}
}

@article{DFT_orbitals_CCSDT_Peterson_JCTC2016,
author = {Fang, Zongtang and Lee, Zachary and Peterson, Kirk A. and Dixon, David A.},
title = {Use of Improved Orbitals for CCSD(T) Calculations for Predicting Heats of Formation of Group IV and Group VI Metal Oxide Monomers and Dimers and UCl6},
journal = {Journal of Chemical Theory and Computation},
volume = {12},
number = {8},
pages = {3583-3592},
year = {2016},
doi = {10.1021/acs.jctc.6b00327},
URL = {https://doi.org/10.1021/acs.jctc.6b00327}
}

@article{steele2009non,
  title={Non-covalent interactions with dual-basis methods: Pairings for augmented basis sets},
  author={Steele, Ryan P and DiStasio Jr, Robert A and Head-Gordon, Martin},
  journal={Journal of Chemical Theory and Computation},
  volume={5},
  number={6},
  pages={1560--1572},
  year={2009},
  publisher={ACS Publications}
}

@article{steele2006dual,
  title={Dual-basis second-order M{\o}ller-Plesset perturbation theory: A reduced-cost reference for correlation calculations},
  author={Steele, Ryan P and DiStasio, Robert A and Shao, Yihan and Kong, Jing and Head-Gordon, Martin},
  journal={The Journal of chemical physics},
  volume={125},
  number={7},
  year={2006},
  publisher={AIP Publishing}
}

@article{steele2007dual,
  title={Dual-basis self-consistent field methods: 6-31G* calculations with a minimal 6-4G primary basis},
  author={Steele, Ryan P and Head-Gordon, Martin},
  journal={Molecular Physics},
  volume={105},
  number={19-22},
  pages={2455--2473},
  year={2007},
  publisher={Taylor \& Francis}
}

@article{liang2004approaching,
  title={Approaching the basis set limit in density functional theory calculations using dual basis sets without diagonalization},
  author={Liang and Head-Gordon, Martin},
  journal={The Journal of Physical Chemistry A},
  volume={108},
  number={15},
  pages={3206--3210},
  year={2004},
  publisher={ACS Publications}
}

@article{hait2019levels,
  title={What levels of coupled cluster theory are appropriate for transition metal systems? A study using near-exact quantum chemical values for 3d transition metal binary compounds},
  author={Hait, Diptarka and Tubman, Norman M and Levine, Daniel S and Whaley, K Birgitta and Head-Gordon, Martin},
  journal={Journal of chemical theory and computation},
  volume={15},
  number={10},
  pages={5370--5385},
  year={2019},
  publisher={ACS Publications}
}

@article{cotton2022truncated,
  title={A truncated Davidson method for the efficient “chemically accurate” calculation of full configuration interaction wavefunctions without any large matrix diagonalization},
  author={Cotton, Stephen J},
  journal={The Journal of Chemical Physics},
  volume={157},
  number={22},
  year={2022},
  publisher={AIP Publishing}
}

@article{wang2025more,
  title={More Numerical Precision for Less Compute Cost: Optimizing a Local Correlation Algorithm for Second Order M{\o}ller--Plesset Theory and Comparing against Pair Natural Orbital Methods},
  author={Wang, Zhenling and Ling, Haobo and Shi, Tianyi and Shen, Yao and Wang, Zhe and Liu, Yang and Li, Xiaoye S and Head-Gordon, Martin},
  journal={Journal of Chemical Theory and Computation},
  volume={21},
  number={21},
  pages={10910--10929},
  year={2025},
  publisher={ACS Publications}
}

\pagebreak
\widetext
\begin{center}
\textbf{\large Supplemental Information}
\end{center}

\setcounter{equation}{0}
\setcounter{figure}{0}
\setcounter{section}{0}
\setcounter{table}{0}
\setcounter{page}{1}
\makeatletter
\renewcommand{\theequation}{S\arabic{equation}}
\renewcommand{\thefigure}{S\arabic{figure}}
\renewcommand{\thetable}{S\Roman{table}}

\section{The effect of density fitted integrals in the  potential energy curve of N$_2$}

\begin{figure}
    \centering
    \includegraphics[width=0.5\textwidth]{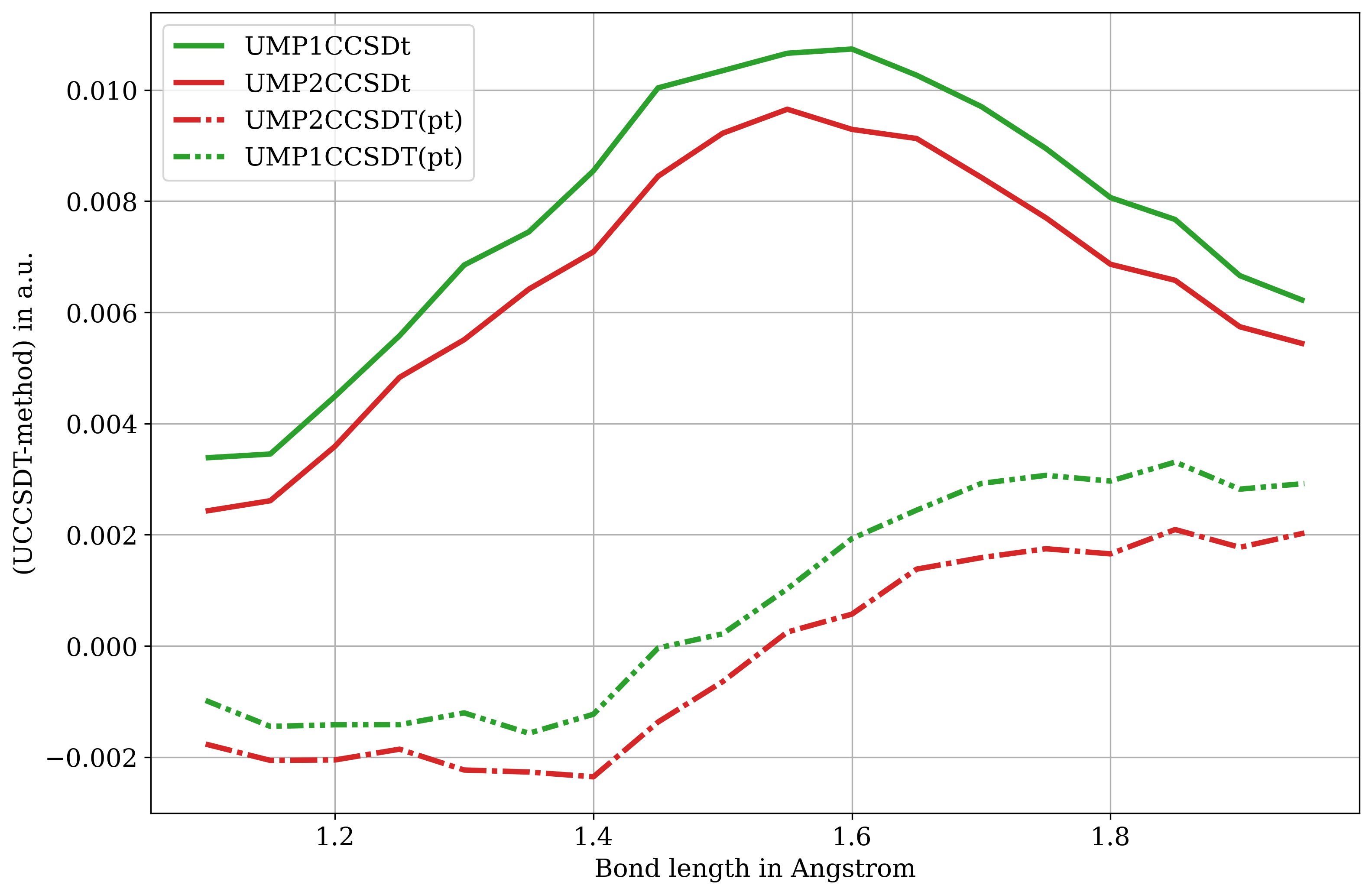}
    \caption{Interaction energy error with respect to UCCSDT along a PEC for N$_2$ in the cc-pCVTZ basis set. AVAS active space is constructed with the cc-pVDZ basis set. Density-fitted integrals are used for all the methods described here.}
    \label{fig:N2_pes_DF}
\end{figure}

We have analyzed the effect of using density-fitted integrals in the study of the potential energy curve of the N$_2$ molecule in Fig. \ref{fig:N2_pes_DF}. We observe that when DF integrals are used for iterative UMP1CCSDt and UMP2CCSDt methods, the error curve w.r.t. the CCSDT method is not smooth. As a result, when the environment triples correction is added to those methods via the UMP1CCSDT(pt) and UMP2CCSDT(pt) methods, the same error curve again appeared to be non-smooth. Therefore, we decided to plot those curves using the non-DF integrals. Furthermore, we observe the same non-smoothness when tried to capture the environment triples corrections iteratively. However, in that case, we avoided performing the same calculation.

\section{Further analysis of the transition metal hydride molecules}

In order to facilitate further analysis of the TMH results presented in Table \ref{Table:ccsdt_tmh}, we have provided two additional tables in this section - one that reports the BDE values at the  CCSD, CCSD(T) and CCSDT level, starting from the ROHF MO basis, and the other showing the total energy values at various levels of theory.

\begin{table}[ht]
\centering
\caption{Total energies (in Hartree) for MnH, Mn, and H obtained with different coupled-cluster methods.}
\label{tab:mnh_energies_rounded}

\begin{tabular}{l
                S[table-format=-4.5]
                S[table-format=-4.5]
                S[table-format=-1.5]}
\hline
Method & MnH & Mn & H \\
\hline
UCCSDT & -1151.19179 & -1150.63583 & -0.49928 \\
UCCSD  & -1151.17691 & -1150.61939 & -0.49928 \\
UCCSD(T) & -1151.19178 & -1150.63603 & -0.49928 \\
UMP1CCSD & -1151.18690 & -1150.63082 & -0.49928 \\
UMP2CCSD & -1151.17085 & -1150.61326 & -0.49928 \\
UMP1CCSDt & -1151.18891 & -1150.63322 & -0.49928 \\
UMP1CCSDT(pt) & -1151.20315 & -1150.64908 & -0.49928 \\
UMP1CCSDT(it) & -1151.19394 & -1150.63932 & -0.49928 \\
UMP2CCSDt & -1151.17281 & -1150.61556 & -0.49928 \\
UMP2CCSDT(pt) & -1151.18595 & -1150.63011 & -0.49928 \\
UMP2CCSDT(it) & -1151.17711 & -1150.62097 & -0.49928 \\
\hline
\end{tabular}
\end{table}

\begin{table}[ht]
\centering
\caption{Total energies (in Hartree) for CrH, Cr, and H obtained with different coupled-cluster methods.}
\label{tab:crh_energies}
\begin{tabular}{l
                S[table-format=-4.5]
                S[table-format=-4.5]
                S[table-format=-1.5]}
\hline
Method & CrH & Cr & H \\
\hline
UCCSDT & -1044.65213 & -1044.07603 & -0.49928 \\
UCCSD  & -1044.63332 & -1044.06079 & -0.49928 \\
UCCSD(T) & -1044.64968 & -1044.07640 & -0.49928 \\
UMP1CCSD & -1044.64420 & -1044.07163 & -0.49928 \\
UMP2CCSD & -1044.62673 & -1044.05417 & -0.49928 \\
UMP1CCSDt & -1044.64745 & -1044.07371 & -0.49928 \\
UMP1CCSDT(pt) & -1044.66471 & -1044.08888 & -0.49928 \\
UMP1CCSDT(it) & -1044.65425 & -1044.07866 & -0.49928 \\
UMP2CCSDt & -1044.63078 & -1044.05628 & -0.49928 \\
UMP2CCSDT(pt) & -1044.64637 & -1044.06990 & -0.49928 \\
UMP2CCSDT(it) & -1044.63650 & -1044.06061 & -0.49928 \\
\hline
\end{tabular}
\end{table}

\begin{table}[ht]
\centering
\caption{Total energies (in Hartree) for FeH, Fe, and H obtained with different coupled-cluster methods.}
\label{tab:feh_energies}
\begin{tabular}{l
                S[table-format=-4.5]
                S[table-format=-4.5]
                S[table-format=-1.5]}
\hline
Method & {FeH} & {Fe} & {H} \\
\hline
UCCSDT & -1263.86904 & -1263.27224 & -0.49928 \\
UCCSD  & -1263.84651 & -1263.25094 & -0.49928 \\
UCCSD(T) & -1263.86392 & -1263.27281 & -0.49928 \\
UMP1CCSD & -1263.86459 & -1263.27438 & -0.49928 \\
UMP2CCSD & -1263.83864 & -1263.24223 & -0.49928 \\
UMP1CCSDt & -1263.86926 & -1263.27769 & -0.49928 \\
UMP1CCSDT(pt) & -1263.88707 & -1263.29939 & -0.49928 \\
UMP1CCSDT(it) & -1263.87520 & -1263.28553 & -0.49928 \\
UMP2CCSDt & -1263.84357 & -1263.24589 & -0.49928 \\
UMP2CCSDT(pt) & -1263.86062 & -1263.26444 & -0.49928 \\
UMP2CCSDT(it) & -1263.84935 & -1263.25288 & -0.49928 \\
\hline
\end{tabular}
\end{table}

\begin{table}[ht]
\centering
\caption{Total energies (in Hartree) for CoH, Co, and H obtained with different coupled-cluster methods.}
\label{tab:coh_energies_rounded}
\begin{tabular}{l
                S[table-format=-4.5]
                S[table-format=-4.5]
                S[table-format=-1.5]}
\hline
Method & {CoH} & {Co} & {H} \\
\hline
UCCSD & -1382.90418 & -1382.31492 & -0.49928 \\
UCCSD(T) & -1382.91924 & -1382.33940 & -0.49928 \\
UCCSDT & -1382.93188 & -1382.33840 & -0.49928 \\
UMP1CCSD & -1382.93165 & -1382.34874 & -0.49928 \\
UMP2CCSD & -1382.89462 & -1382.30480 & -0.49928 \\
UMP1CCSDt & -1382.93764 & -1382.35269 & -0.49928 \\
UMP1CCSDT(pt) & -1382.95806 & -1382.37716 & -0.49928 \\
UMP1CCSDT(it) & -1382.94414 & -1382.35636 & -0.49928 \\
UMP2CCSDt & -1382.90103 & -1382.30933 & -0.49928 \\
UMP2CCSDT(pt) & -1382.92085 &	-1382.32966	& -0.49928\\
UMP2CCSDT(it) & -1382.90985 & -1382.31722 & -0.49928 \\
\hline
\end{tabular}
\end{table}

\begin{table}[ht]
\centering
\caption{Bond dissociation energies (in kJ/mol) for MnH, CrH, CoH and FeH obtained with RO-based coupled-cluster methods.}
\label{tab:bde_ro_methods}
\begin{tabular}{lcccc}
\hline
Method & {MnH} & {CrH} & {CoH} & {FeH} \\
\hline
ROCCSD    & 153.11 & 191.66 & 233.71 & 241.54 \\
ROCCSD(T) & 148.47 & 191.36 & 245.52 & 251.76 \\
ROCCSDT   & 151.98 & 191.36 & 245.17 & 247.59 \\
\hline
\end{tabular}
\end{table}

\end{document}